\documentclass{article} 
\usepackage[sumlimits]{amsmath}
\usepackage{paralist,amssymb,siunitx,gensymb}
\usepackage{bm}
\usepackage[breaklinks=true]{hyperref}
\usepackage{url, natbib, cite}
\usepackage{breakcites}
\usepackage{graphicx}
\usepackage{rotating}
\usepackage{authblk}
\DeclareMathOperator*{\argmin}{arg\,min}

\begin{document}
\title{Ultrahigh Dimensional Variable Selection for Mapping Soil Carbon}
\author[1,2,3,*]{Benjamin R. Fitzpatrick}
\author[4,2]{David W. Lamb}
\author[5,1,2,3]{Kerrie Mengersen}
\affil[1]{\small Mathematical Sciences School, Queensland University of Technology (QUT), Brisbane, QLD 4001, Australia}
\affil[2]{\small Cooperative Research Centre for Spatial Information (CRCSI), Carlton, VIC 3053, Australia} 
\affil[3]{\small Institute for Future Environments, Queensland University of Technology (QUT), Brisbane, QLD 4001, Australia}
\affil[4]{\small Precision Agriculture Research Group, University of New England, Armidale, NSW 2351, Australia}
\affil[5]{\small ARC Centre of Excellence for Mathematical and Statistical Frontiers, Queensland University of Technology (QUT), Brisbane, QLD 4001, Australia}
\affil[*]{\small E-mail: Corresponding \texttt{b1.fitzpatrick@qut.edu.au}}
\maketitle

\begin{abstract} 
Modern soil mapping is characterised by the need to interpolate samples of geostatistical response observations and the availability of relatively large numbers of environmental characteristics for consideration as covariates to aid this interpolation.
We demonstrate the efficiency of the Least Angle Regression algorithm for Least Absolute Shrinkage and Selection Operator (LASSO) penalized multiple linear regression at selecting covariates to aid the spatial interpolation of geostatistical soil carbon observations under an ultrahigh dimensional scenario.  
Where an exhaustive search of the models that could be constructed from 800 potential covariate terms and 60 observations would be prohibitively demanding, LASSO variable selection is accomplished with trivial computational investment.
\end{abstract}

\section{Introduction}
Global soils have been estimated to contain the largest pool of terrestrial organic carbon in the biosphere, storing more carbon than all land plants and the atmosphere combined \citep{Schlesinger1977}.
The importance of the dynamic equilibrium between carbon in soils and carbon in the atmosphere has been illustrated by such estimates as there having been 3.3 times the amount of carbon in the atmosphere as $CO_2$(g) present in global soils \citep{Lal2004}.
More than half of the global soil carbon pool has been estimated to be comprised of organic compounds collectively referred to as soil organic carbon (hereafter SOC) \citep{Lal2004}. 
SOC may be depleted to as little as 25\% of capacity when natural ecosystems are converted into agricultural systems with the majority of this carbon lost to the atmosphere as $CO_2$(g) \citep{Lal2004}.
The contribution such SOC losses would have made to terrestrial carbon dynamics may be appreciated in the context of the estimate that 34\% of the global land surface had been devoted to agriculture by 2007 \citep{Betts2007}.
Recharging SOC levels by sequestering $CO_2$(g) in agricultural soils has been demonstrated to provide direct benefits to agriculture, in addition to providing an opportunity to partially offset anthropogenic green house gas emissions \citep{Lal2009}.  
Consequently, it is a key feature of national and international level carbon accounting endeavours.
\newline
\newline 
The effort and cost associated with sampling SOC via laboratory analysis of soil core samples has led to a need to improve soil core sample based maps of SOC through statistical modelling using more readily attainable environmental variables as covariates \citep{Mueller2003, Barnes2003, Chan2008, Johnson2001, Simbahan2006b, Miklos2010, Vasques2012}.  
Such modelling is often accompanied by two challenges.
The first is spatial misalignment of observations of different variables and or observations and the locations (or coverage extents and resolutions) to which SOC is to be interpolated.
The second is the availability of large numbers of potentially relevant covariates coupled with the belief that selected model(s) should be sparse.
In this paper, we address the challenges of spatial misalignment and selection of a parsimonious subset of covariates to aid spatial interpolation of the response under an ultrahigh dimensional scenario.
We achieve this by showcasing the performance of Least Absolute Shrinkage Selection Operator (LASSO) penalized Multiple Linear Regression (MLR) models on data from a real world case study of soil core derived observations of \%SOC across 137ha of agricultural land in New South Wales, Australia.
The remainder of the article is structured as follows.
Section 2 describes the field sites along with the data collection, collation and spatial realignment for our case study.
In Section 3 we explain the motivation of our choice of LASSO variable selection and summarize the key characteristics of this method.
Section 4 contains the results and discussion of our analysis of the case study data.
In particular we compare LASSO variable selection to four popular variable selection methods, evaluate the set of selected covariates and describe fitting spatial polynomials to the residuals for more precise interpolation of \%SOC.
We describe how we combine the predictions from these spatial polynomials with the predictions from the covariate based modelling to produce a full cover predicted raster for \%SOC.
We conclude with a discussion of this work and promising avenues for future research in Section 5.

\section{Data Collection \& Preparation}

\subsection{Data Collection \& Collation}
\label{sec:site.data}
Our case study data were collected from a 137ha area of native pasture with remnant woody vegetation on the Sustainable, Manageable, Accessible, Rural Technology (SMART) Farm of the University of New England near Armidale, New South Wales, Australia.
The 60 observations of our response variable, percentage soil organic carbon (\%SOC), include 57 values less than 2.55\% while the remaining three values are 3.08\%, 5.01\% and 5.13\%.  
We summarize the 63 environmental characteristics we consider here as potential covariates in Table 1.
The Digital Elevation Model (DEM) derived covariates were calculated with the System for Automated Geoscientific Analyses (SAGA v2.1.0 \citep{Conrad2015}) software and the resulting GIS layers were read into R \citep{R2015} with the `RSAGA' \citep{RSAGA2015} package.  
The remaining raster covariates were read into R with the R package `raster' \citep{Hijmans2012}.
Further details regarding the study site, field methodology and covariates are provided in Appendices A and B.

\begin{table}
\caption{The 63 potential covariates.}
\begin{tabular}{ p{3.5cm} p{7cm} p{2cm} }
Source & Covariate Name & Acronym \\
\hline
\hline
ATV Top of Pasture & Soil Apparent Electrical Conductivity & ECA \\
Surveys & Near InfraRed Reflectance & NIR \\
12 covariates & Red Reflectance & RED \\
from each of February, & Simple Ratio & SR \\
May \& November & Difference Vegetation Index & DVI \\
 = 36 covariates & Normalized Difference Vegetation Index & NDVI \\
 & Soil Adjusted Vegetation Index & SAVI \\
 & Non-Linear Vegetation Index & NLVI \\
 & Modified Non-Linear Vegetation Index & MNLVI \\
 & Modified Simple Ratio & MSR \\
 & Transformed Vegetation Index & TVI \\
 & Re-normalised Difference Vegetation Index & RDVI \\
\hline
Terrain \&& Catchment Area & CatAr \\
Hyrdology Metrics & Catchment Height & CatHe \\
Calculated from & Catchment Slope & CatSl\\
$25m^2$ resolution & Cosine(Aspect) & CosAsp \\
DEM & Elevation &  Elev \\
= 16 Covariates & Slope Length Factor & LSF \\
& Plan Curvature & PlanC \\
& Profile Curvature & ProfC \\
& Sky View Factor & SVF \\
& Slope & Slp \\
& Stream Power Index & SPI \\
& Terrain Ruggedness Index & TRI \\
& Topographic Position Index & TPI \\
& Vector Terrain Ruggedness & VTR \\
& Visible Sky & VS \\
& Wetness Index & WI \\
\hline
Foliar Projective Cover Layers & 2011 & FPCI\\
= 2 Covariates & 2012 & FPCII \\
\hline
Electromagnetic Channels & 1 to 6 & MagI - MagVI \\
 = 6 Covariates &  &  \\
\hline
$\gamma$ Radiometric Layers & Potassium & K \\
= 3 Covariates & Thorium & Th \\
 & Uranium & U \\
\hline
\hline
\end{tabular}
\end{table}

\subsection{Spatial Realignment of Covariate and Response Observations} 
\label{sec:prep.of.data}
The geostatistical response observations are spatially misaligned with the geostatistical covariate observations and with the raster covariate observations.
As the majority of our raster based covariates are derived from the 25m$^2$ resolution DEM, we realign all covariates to 25m by 25m squares centered on each response observation.
Geostatistical covariates are interpolated to regular 100 by 100 point grids centred on the response observations via thin plate splines with the R package `fields' \citep{Nychka2015}.
The values of raster covariates are queried at these same grids of points centred on each response observation.
The realigned value of each covariate associated with each response observation is taken as the mean of the values of this covariate across the grid of points centred on that response observation.

\section{Statistical Background}
\subsection{Choice of Modelling Method}
\label{sec:MLR.cf.BT}
A variety of statistical and machine learning techniques have been applied to soil carbon modelling.
Such techniques include ANOVA \citep{Johnson2001}, multiple linear regression (MLR) \citep{Mao2015}, MLR with stepwise variable selection \citep{Moore1993, Mueller2003, Terra2004, Florinsky2002, Meersmans2012}, MLR on the principal components of the covariate observations \citep{Wiesmeier2013}, regression fitted by partial least squares \citep{Hbirkou2012a}, MLR with stepwise variable selection within groups of the data identified via neural networks \citep{Chen2008} and regression kriging \citep{Dlugoss2010, Simbahan2006b}.
Binary tree based methods applied to soil carbon modeling include Classification And Regression Trees \citep{BouKheir2010, Wiesmeier2010}, Random Forests \citep{Wiesmeier2010} and CUBIST \citep{Miklos2010, Lacoste2014, Adhikari2014, Xiong2014, ViscarraRossel2014}.
We evaluate the advantages and disadvantages of a range of models in terms of our objective (covariate assisted spatial interpolation), computational demands and the three defining characteristics of our case study data: (1) more potential covariate terms than observations (ultrahigh dimensionality) (2) a high degree of collinearity among the potential covariate terms and (3) suspected importance of non-linear effects of covariates and interactions of covariate effects.
The MLR based approaches we consider include: ridge regression \citep{Hoerl1970}, Least Absolute Shrinkage and Selection Operator (hereafter LASSO) modified MLR fitted via quadratic programming \citep{Tibshirani1996}, LASSO modified MLR fitted by the Least Angle Regression (hereafter LAR) algorithm \citep{Efron2004} and the Bayesian LASSO \citep{Park2008}.
The CART based techniques we consider include: Bayesian CART \citep{Chipman1998}, bagged regression trees \citep{Breiman1996}, random forests \citep{Breiman2001}, boruta all relevant variable selection \citep{Kursa2010}, boosted regression trees \citep{Friedman2002}, cubist \citep{Quinlan1992} (\url{https://www.rulequest.com/cubist-info.html}) and Bayesian treed regression \citep{Chipman2002}.
This evaluation is summarised in Appendix C.
\newline
\newline
We use LASSO modified MLR fitted via the LAR algorithm in our case study analysis. 
Model-averaging the predictions from the LASSO solutions obtained from LAR executions within a cross validation scheme yields an aggregate estimate in a manner similar to random forests, bagged trees and boosted trees.
A cross validation based approach also facilitates estimation of the shrinkage parameter for the LASSO fits ($\lambda$ in Equation \ref{eq:L.gamma.pen.ls}).
The choice of LASSO modified MLR allows the importance of covariate terms (linear, non-linear and interaction) to be compared in terms of which have the coefficients that are shrunk to zero and which are assigned non-zero values.
In contrast, whether the overall role of a covariate within the aggregated estimate from random forests, bagged or boosted trees is closer to linear or non-linear (and if non-linear what manner of non-linear) would be harder to judge from the results of such a fit.
This ease of interpretability of the LASSO modified MLR comes with the cost of having to recenter and rescale (to mean zero and magnitude one) all covariates in each training each set (a requirement of the LAR algorithm \citep{Efron2004}) and mirror those transformations on each associated validation set.
Whereas, such transformations are unnecessary for binary tree based techniques. 

\subsection{LASSO Variable Selection as a Special Case of PLS} 
\label{sec:VS_MA}
Penalized Least Squares (PLS) coefficient estimates ($\bm{\hat \beta}$ in Equation \ref{eq:L.gamma.pen.ls}) are calculated by identifying the coefficient estimate vector that minimizes the sum of the residual sum of squares and the result of applying some penalty function to the coefficients.
Simple PLS estimates use the $L_\gamma$ norm $ \sum\limits_{j = 1}^p | \beta_j |^\gamma$ of the coefficient vector $\bm \beta$ for some $\gamma > 0$ as the penalty function so that
\begin{equation} \label{eq:L.gamma.pen.ls}
\bm{\hat \beta} = \argmin\limits_{\bm{\beta}} \{ \sum\limits_{i = 1}^n (y_{i} - \beta_0 - \sum\limits_{j = 1}^px_{ij} \beta_j)^2 + \lambda \sum\limits_{j = 1}^p | \beta_j |^\gamma  \},\ \ \gamma > 0
\end{equation}
where the tuning parameter $\lambda$ controls the degree to which $\bm{\hat \beta}$ is shrunk towards the zero vector \citep{Ahmed2014}.
When $\gamma$ is set to 1, the solution to Equation \ref{eq:L.gamma.pen.ls} is the $L_1$ PLS estimate of $\bm{\beta}$, also known as the Least Absolute Shrinkage and Selection Operator (LASSO) \citep{Tibshirani1996}.
When $\gamma$ is set to 2, the solution to Equation \ref{eq:L.gamma.pen.ls} is the $L_2$ PLS estimate of $\bm{\beta}$ which is referred to as a ridge regression estimate \citep{Hoerl1970}.
Other penalized least squares techniques including adaptive LASSO \citep{Zou2006}, Smoothly Clipped Absolute Deviation (SCAD) \citep{Fan2001} and Minimax Concave Penalty (MCP) \citep{Zhang2010} are derived through use of more complex penalty functions in place of the $L_\gamma$ norm in Equation \ref{eq:L.gamma.pen.ls}.
Solving Equation \ref{eq:L.gamma.pen.ls} with $\gamma$ set to a value of 2 or less results in the values of some coefficients being estimated as zero exactly (how many depends on the value of the tuning parameter $\lambda$) \citep{Ahmed2014}.
Since a coefficient estimate of zero is equivalent to exclusion from the selected model such a solution effectively performs both variable selection and shrinkage.
As such, $L_\gamma$ penalized estimation with $\gamma < 2$ is applicable to our case study where the number of potential covariates exceeds the number of observations ($p > n$). 
\newline
\newline
The requirement for a computational solution to $L_1$ penalized estimation (stemming from the presence of the absolute value in Equation \ref{eq:L.gamma.pen.ls}) was originally addressed via relatively computationally expensive quadratic programming \citep{Tibshirani1996} and has been addressed more recently by the computationally efficient Least Angle Regression (LAR) algorithm \citep{Efron2004}.
From the PLS family of techniques we chose to adopt $L_1$ penalized estimation for three reasons: \begin{inparaenum}[1)] \item suitability for variable selection and modelling with correlated covariates \item suitability for variable selection in scenarios with $p > n$ and \item the computational efficiency of the LAR algorithm \citep{Efron2004}\end{inparaenum}.
\newline
\newline
The LAR algorithm has been designed such that covariates continue to be added to the model until either the available degrees of freedom are exhausted or there are no covariates outside the current model that have a correlation with the current residual vector greater in magnitude than some user specified threshold value.
In the case of the LASSO modification of the LAR algorithm, while steps of the algorithm may result in a covariate being removed from the current model, the algorithm still proceeds to add and remove covariates from the current model until either of the above criteria are met.
Subsequently, the LAR algorithm (and the LASSO variant thereof) returns a sequence of selected models from which it is necessary to choose a parsimonious final model.
Efron et al. [2004] derive a $C_p$ style stopping criterion for the LAR algorithm but note that this is most appropriate in scenarios with less potential covariates than observations.  
Alternative stopping criteria, applicable to more general scenarios, also exist \citep{Valdman2012} though cross validation is a popular approach for ultrahigh dimensional problems \citep{Engelmann2012, Usai2012, Usai2009}.
Hence, a cross validation based approach to making the final selection from the sequence of selected models produced by the LAR algorithm is adopted here.
All analysis is conducted in the R language and environment for statistical computing \citep{R2015} and all graphics are produced with the R package `ggplot2' \citep{Wickham2009}.
The data and R code associated with this work will be provided via a repository located at \url{https://github.com/brfitzpatrick/larc} once this work is published in a peer reviewed journal.

\section{Methods and Results}
\subsection{Comparison of Variable Selection Methods for MLR} \label{sec:TSS.VSMC}
We compare LASSO variable selection to the more generic variable selection methods: exhaustive search, forward stepwise selection, backwards stepwise selection and sequential replacement selection (also known as stepwise forwards-backwards variable selection) on the case study data.
Due to the complexity of interacting processes that may influence the formation, distribution and loss of SOC across the study site we consider polynomial terms up to order four for each covariate and all possible pairwise interactions of the covariates.
The full set of potential covariates thus expands from 63 to 2205 potential covariate terms ($63*4 + \binom{63}{2}$).
With 60 observations of the response, if we wished to explore all possible models from an intercept only model up to those that used the available degrees of freedom, we would need to fit and compare some $ \sum\limits_{i = 1}^{60} \binom{2206}{i} \approx 2.27*10^{118} $ different models in an exhaustive search.
To reduce the number of covariates considered and thus the required breadth of exhaustive search we pre-filter the design matrix such that no remaining pair of covariates have a correlation coefficient greater in magnitude than some critical value.
Since the correlation of a potential covariate with the response may be a poor indicator of the explanatory utility of this covariate in the presence of other covariates, we choose between highly correlated pairs of covariates based upon the spatial resolution at which each is available.
The motivation behind this decision being an effort to optimise the spatial accuracy of our interpolation of the response.
For covariates with the same spatial resolution, the one derived from the simpler function of observed data is chosen, otherwise the choice is made at random.
These criteria are discussed in more detail in Appendix D.
\newline
\newline
Pre-filtering to enforce a maximum correlation coefficient magnitude (hereafter MCCM) of 0.4 between remaining covariate pairs results in a design matrix with 27 covariate terms.
The branch-and-bound algorithm implemented in the `leaps' package \citep{Lumley2009} requires only a subset of the $ \sum\limits_{i = 1}^{28}  \binom{28}{i} \approx 2.68*10^8 $ models it is possible to construct from this design matrix to be fitted in order to determine the optimal model that would be returned from a full exhaustive search \citep{Millar2002}.
As our aim here is to build models that make the best use of covariates for spatial interpolation of the response, the metric by which we compare the results of these variable selection techniques is the ability of the selected models to predict data held out from the fitting process. 
We conduct these comparisons on 500 unique divisions of the data into training and validation sets in a cross validation scheme.
We use training and validation sets of 35 and 25 observations, respectively.
The selection of a training set size is discussed in Appendix E.
\newline
\newline
Training sets constructed from the design matrix composed of 27 covariate terms are supplied to each of the variable selection methods (LASSO variable selection, forward selection, backward selection, sequential replacement and exhaustive search variable selection).
In each case the final selection from the sequence of models returned is made to minimise the validation set element prediction error (here after VSEPE) sum of squares.
The distributions of VSEPE absolute values from each variable selection technique are summarized in Table 2.
\begin{table}
\caption{Summary statistics for the absolute values of validation set element prediction error (VSEPE) distributions from each variable selection method conducted on design matrices pre-filtered to enforce a maximum correlation coefficient magnitude between covariate pairs of 0.4 or 0.95 ($|r| \leq 0.4$ or $|r| \leq 0.95$). The final column contains the coefficient of determination ($R^2$) values for the model averaged predictions (MAP) from the models resulting from the combinations of variable selection technique and design matrix pre-filtering austerity specified by that row. LAR = Least Angle Regression Variable Selection, Exh = Exhaustive Search Variable Selection, Seq = Sequential Replacement Variable Selection, Fwd = Forward Stepwise Variable Selection, Bwd = Backward Stepwise Variable Selection, Min. = Minimum , 1st Qu. = First Quartile, 3rd Qu. = Third Quartile and Max. = Maximum.}
\small
\begin{tabular}{lrrrrrrrr}
  \hline
       &         & \multicolumn{6}{ c }{ VSEPE} & MAP \\
Method & $|r|\leq$ & Min.      & 1st Qu. & Median & Mean & 3rd Qu.  & Max.  & $R^2$ \\ 
  \hline
  LAR  &   0.95  & 1.332e-05 & 0.1482  & 0.3184 & 0.4744 & 0.5446 & 4.437 & 0.5963 \\ 
  LAR  &   0.40  & 1.097e-05 & 0.1517  & 0.3324 & 0.4776 & 0.5695 & 4.063 & 0.3666 \\ 
  Exh  &   0.40  & 5.571e-05 & 0.1644  & 0.3419 & 0.4964 & 0.5997 & 4.290 & 0.2882 \\ 
  Seq  &   0.40  & 5.571e-05 & 0.1677  & 0.3448 & 0.4960 & 0.6044 & 3.961 & 0.3055 \\ 
  Fwd  &   0.40  & 5.571e-05 & 0.1604  & 0.3392 & 0.4955 & 0.5994 & 4.063 & 0.3046 \\ 
  Bwd  &   0.40  & 1.036e-05 & 0.1654  & 0.3593 & 0.5053 & 0.6037 & 4.422 & 0.2382 \\ 
   \hline
   \hline
\end{tabular}
\end{table}
When applied to these austerely pre-filtered design matrices, all five variable selection techniques yield very similar VSEPE distributions.
The first three quarters of the ordered VSEPE absolute values obtained from LASSO variable selection are slightly more compressed towards zero than those from any other technique considered.
\newline
\newline
Predictions from the 500 selected models (one per training set) are model-averaged with weights inversely proportional to the prediction error sum of squares on the associated validation sets.
Taking $i$ to index the 500 divisions of the data into training and validation sets, the weights for model-averaging, $W_i$, are calculated following Equation \ref{eq:MA.Weights}.                
Here $e_{i,\,j}$ is the prediction error of the $j^\text{th}$ element of the $i^\text{th}$ validation set where each validation set has $v$ elements.
\begin{equation} \label{eq:MA.Weights}
W_i = \frac{\frac{1}{\sum\limits_{j=1}^{v}{e_{i,\,j}^2}}}{\sum\limits_{i=1}^{500}{\frac{1}{\sum\limits_{j=1}^{v}{e_{i,\,j}^2}}}}
\end{equation}
The noticeable improvement in accuracy of the model-averaged predictions from the models selected by LAR is shown in the column of coefficient of determination values in Table 2.
Corresponding summary statistics for the absolute values of the VSEPE obtained from model fitted to 800 term design matrices that result from using a much less stringent MCCM of 0.95 are also included in Table 2 along with the coefficient of determination for the associated model-averaged predictions.
Similar, but greater magnitude, improvements are observed between the LAR selected models for the 27 covariates design matrices and the 800 covariates design matrices as were observed between models selected by other variable selection techniques and LAR selected models.
These improvements come with an increased computational cost, but the application of the LAR algorithm to these expanded design matrices is still feasible even on a mid range laptop computer whereas exhaustive search variable selection on these expanded design matrices would be infeasible.
The positive outliers in all the VSEPE distributions are likely the result of the three positive outliers in the response.
When these are drawn as members of a validation set, models built from the associated training set likely under-predict these values in the validation set.
\newline
\newline
The distributions of the numbers of covariates selected by each of the variable selection methods from the 27 covariate design matrices are depicted in Figure 1.
\begin{figure}
\centering
\makebox{\includegraphics[height = 0.3\textheight]{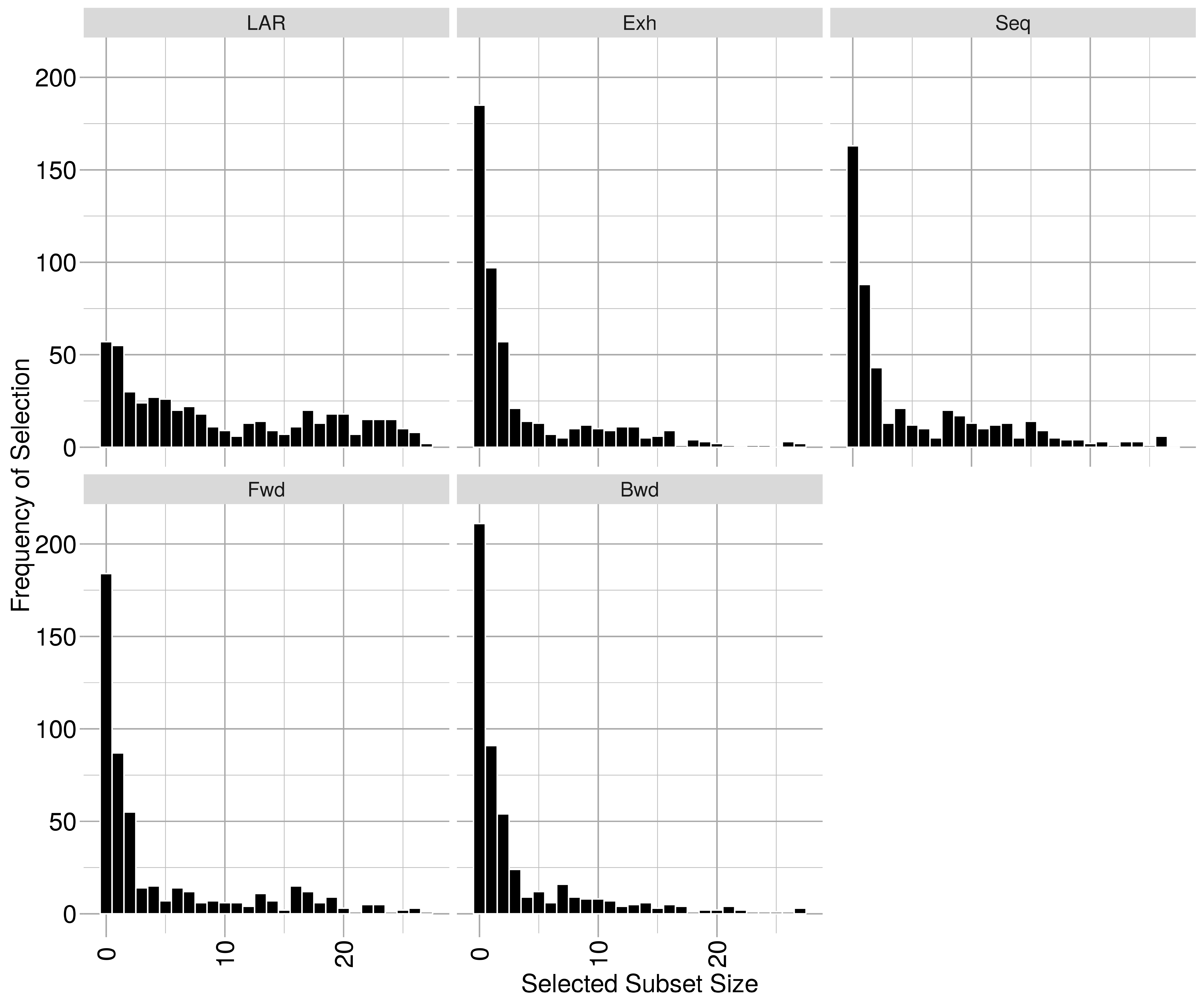}}
\caption{Histograms depicting the distribution of subset sizes selected by each variable selection technique applied to training sets constructed from the 27 covariate design matrix. LAR = Least Angle Regression Variable Selection, Exh = Exhaustive Search Variable Selection, Seq = Sequential Replacement Variable Selection, Fwd = Forward Stepwise Variable Selection, Bwd = Backward Stepwise Variable Selection, Min. = Minimum , 1st Qu. = First Quartile, 3rd Qu. = Third Quartile and Max. = Maximum.}
\end{figure}
The LASSO method results in intercept only models far less frequently and larger numbers of covariates per model more frequently than the other techniques. 
The differences in predictive accuracy and numbers of covariates selected per model, between the LASSO and the forwards stepwise OLS based method may be explained in terms of the comparative theoretical properties of these algorithms.
At each step in the respective algorithms both approaches choose the covariate most correlated with the current residual vector for inclusion in the current model.
However, LAR adds this new covariate to the model in such a manner that the resulting prediction vector is equiangular between the previous prediction vector and this new covariate vector and only proceeds along this new prediction vector until some other covariate outside the current model is as correlated with the current residual vector as the most recently added covariate before repeating this procedure. 
Forwards selection, backwards stepwise variable selection and sequential replacement variable selection lack this facility to compromise between the correlated covariates.
Furthermore, the differences between the results of LASSO variable selection and the exhaustive search variable selection may well stem from exhaustive search variable selection using OLS model fitting while the LASSO variable selection uses PLS based model fitting.

\subsection{Frequently Selected Covariates}
\label{sec:FSC}
The numbers of the 500 selected models in which particular covariate terms occur can serve as an indicator of the relative importance of these terms for predicting the response.
In Table 3 we list the 15 most frequently selected terms from LAR variable selection on the 800 column design matrices.
Table 3 also lists covariate terms from the 2205 column design matrix which were very highly correlated ($|r|>0.95$) with these top 15 covariates and were thus excluded from the analysis.
\begin{table}
\caption{The 15 most frequently selected covariates from the LAR variable selection executions on the 500 unique, 35 observation training sets constructed from the design matrix created by pre-filtering the full design matrix to enforce a maximum permitted correlation coefficient magnitude between remaining covariates pairs of 0.95.  The second column contains the frequencies with which the selected covariates occurred in the 500 selected models.  Accompanying each selected covariate in the final column are the covariates from the full design matrix that had correlation coefficient magnitudes with the covariate in question greater than 0.95 and thus were excluded from the design matrix supplied to the variable selection.  Colons denote interaction terms for the two covariate terms the colon separates.  Numeric superscripts denote polynomial terms for the covariate indicated by the acronym. Acronyms are expanded in Table 1.}
\label{Tab:T20Cov} 
\centering
\begin{tabular}{p{2.75cm} p{0.5cm} p{7cm}}
\hline
Covariate       & Freq & Correlated Covariates                    \\
\hline
\hline
ECA.Nov$^4$     & 219 & -                                          \\
LSF$^3$         & 139 & Slp$^3$, TRI$^3$, LSF$^4$, Slp$^4$, TRI$^4$ \\
DVI.May         & 102 & SAVI.May, NLVI.May, MNLVI.May, RDVI.May    \\
WI              & 100 & -                                          \\
ECA.Feb:Slp     &  95 & ECA.Feb:TRI                                \\
Mag.II:FPCI     &  95 & -                                          \\
SVF:Mag.IV      &  94 & -                                          \\
Slp$^2$         &  89 & LSF:Slp, LSF:TRI, Slp:TRI, TRI:WI, TRI$^2$ \\
ECA.Feb:SR.May  &  88 & ECA.Feb:NDVI.May, ECA.Feb:SAVI.May, ECA.Feb:MSR.May, ECA.Feb:TVI.May, ECA.Feb:RDVI.May \\
LSF:SVF         &  82 & LSF:VTR, SVF:Slp, SVF:TRI                  \\
ECA.Nov:DVI.Nov &  78 & ECA.Nov:MNLVI.Nov                          \\
Elev:SVF        &  76 & -                                          \\
ECA.Feb:DVI.Nov &  74 & ECA.Feb:MNLVI.Nov, ECA.Feb:RDVI.Nov        \\
ECA.Nov$^3$     &  73 & -                                          \\
ECA.Feb:Elev    &  72 & -                                          \\
\hline
\hline
\end{tabular}
\end{table}
A chord diagram depicting the selection frequencies of all 800 covariate terms is presented in Figure 2.
\begin{figure}
\centering
\makebox{\includegraphics[height = 0.35\textheight]{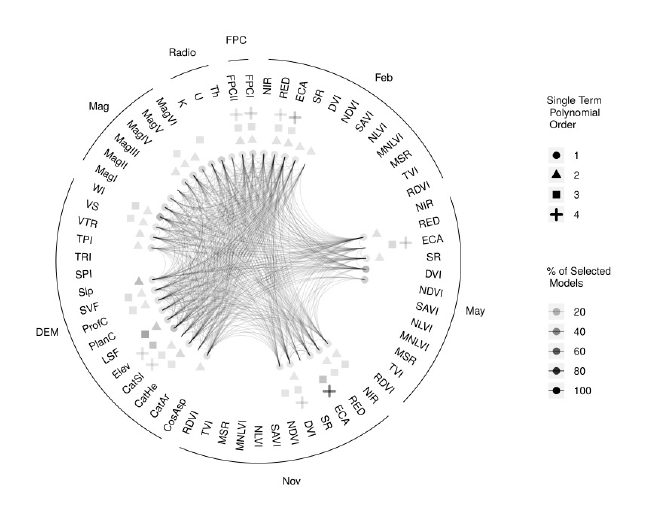}}
\caption{Covariate term selection frequencies across the 500 selected models obtained from applying the Least Angle Regression variable selection algorithm to the 35 observation training sets constructed from design matrices produced by pre-filtering the full design matrix to enforce a maximum permitted correlation coefficient magnitude between covariate pairs of 0.95. Poincar\'{e} segments represent interaction terms between the covariates they connect. Covariate Acronyms are expanded in Table 1.}
\end{figure}
The complexity of interacting processes producing the spatial distributions of SOC in agricultural landscapes like that of our case study site is reflected in the diversity of the categories of covariates terms selected (soil $EC_a$, vegetation indices, DEM derived metrics, magnetic imagery, radiometric imagery and foliar projective cover layers) and the mixture of linear terms, higher order polynomial terms and interactions of linear terms selected for these covariates.

\subsection{Modelling Spatial Component of Error}
\label{ss:spat.comp.err}
Following the model-averaging described above we fit a spatial model to the residual \%SOC variation at each soil core location.
This allows spatial position to serve as a locally appropriate proxy for all the unobserved processes and interactions that may influence the spatial distribution of \%SOC at our case study site.
One approach would be to use Kriging to spatially interpolate the residuals, but this requires the comparison of numerous  pairs of orthogonal, directional, empirical semivariograms.
A more attractive alternative is to calculate an empirical semivariogram raster, in which pairwise differences between geostatistical observations are assigned to two dimensional displacement bins and the empirical semivariance is calculated for each bin.
The resulting raster may then either be smoothed \citep{Banerjee2003} or simply examined directly and the spatial symmetry of the resulting values considered.
In our study, however the small sample size would result in moderate numbers of pairs per bin only when a relatively large bin size is used.
The resulting coarse spatial resolution would make characterisation any detected anisotropy infeasible.
For this reason we adopt the simpler approach of fitting spatial polynomials to the residuals and model-averaging the results via the same procedure we use for the covariate based modelling.
\newline
\newline
The computational efficiency of the LAR algorithm enables us to explore design matrices that include single term polynomials for Easting and Northing values up to polynomial order 12 and interactions terms constructed from subsets of these single term polynomials such that all possible product terms which equate to an overall polynomial order of 6 or less are included in this exploration.
We only consider interaction terms equivalent to a polynomial term of half the order of the maximum order of polynomial terms considered in order to avoid confounding between interactions terms of order equivalent to the higher order single polynomial terms.
We use the results of fitting the spatial polynomials to training sets of 35 observations constructed from the design matrix pre-filtered to enforce a MCCM between covariate pairs of 0.95 for similar reasons involved in this decision for the covariate based variable selection.
Again, 500 unique divisions of the data into training and validation sets are constructed and explored by LAR variable selection and final selections are made from each LAR model choice trajectory on the basis of which model minimizes the associated VSEPE sum of squares.
Model-averaging is conducted with weights inversely proportional to the VSEPE sums of squares as per Equation \ref{eq:MA.Weights}.

\subsection{Full Cover Inference}
The 500 selected models (each selected for one of the unique training sets) yield 500 predicted values for \%SOC at every pixel in the final prediction raster.
We use the weighted model-averaging procedure described in Section 4.1 to calculate a \%SOC prediction for each of these pixels.
We also calculate an uncertainty estimate for these predictions, where the uncertainty is quantified by the width of the interval containing the middle 95\% of the predictions for that pixel.
A panel of two rasters is presented in Figure 3.
The areal prediction of \%SOC levels across the study area plus the areal prediction of the spatial component of the errors from the covariate based modelling is presented as the top raster in Figure 3.
The predictions for each pixel from the covariate based modelling are constructed by model-averaging the predictions for that pixel from the models selected by LAR exploration of the 500 unique 35 observation training sets constructed by subsetting the 800 column design matrix.
Our estimate of the uncertainty associated with these predictions is presented as the bottom raster in Figure 3.
The predicted spatial distribution of \%SOC levels is overall quite uniform across the study site with only a few localized regions of notably elevated or depressed values.
The estimated uncertainty associated with the predicted \%SOC levels is relatively low across the majority of the study site with a few regions of notably elevated uncertainty.

\begin{figure}
\centering
\makebox{\includegraphics[height = 0.75\textheight]{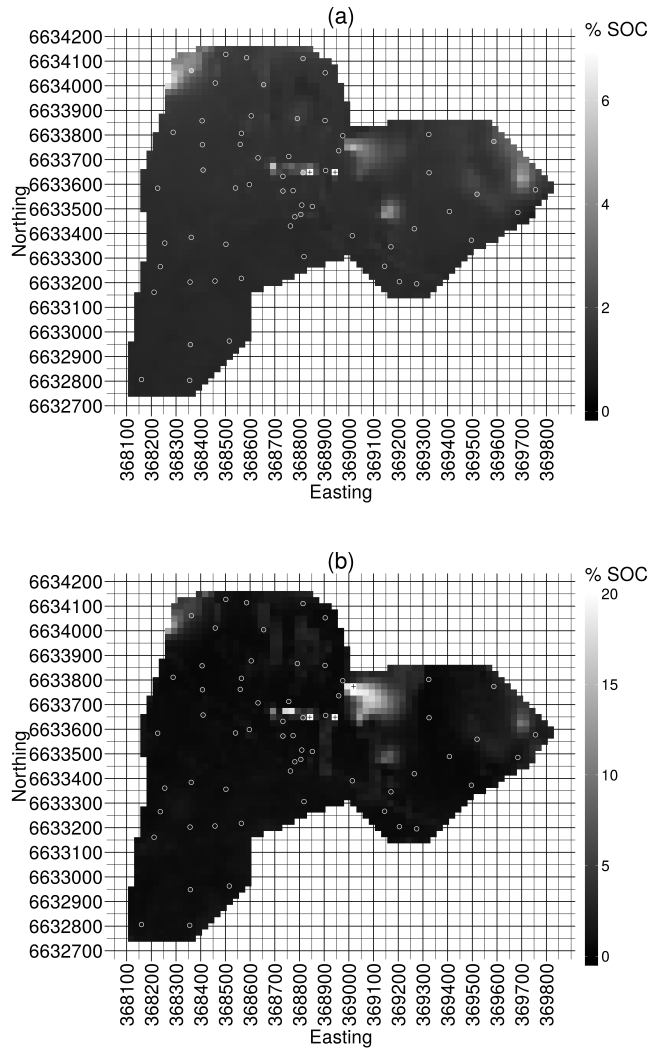}}
\caption{The observed soil organic carbon percentages (\%SOC) at the soil core locations have been represented by the shade filling the circles located at each of the soil core sample locations. 
The observed \%SOC values have been represented with the same grey scale as the predicted \%SOC values and associated uncertainties in the rasters.
\textbf{(a)} The sum of the covariate based predictions and the predictions from the model for the spatial component of the errors from the covariate based model. The more westerly pixel annotated with a vertical cross represents a predicted \%SOC value of 17.92 and the more easterly pixel annotated with a vertical cross represents a predicted \%SOC value of 9.54.
\textbf{(b)} The uncertainty estimated to accompany the \%SOC predictions.
The three pixels annotated with vertical crosses represent estimates of the uncertainty associated with the model-averaged predicted \%SOC values of 20.57, 21.66 and 43.66 units on the predicted \%SOC scale.
The estimated uncertainty of 43.66 being the most westerly of these three pixels and the estimated uncertainty of 20.57 being the most northerly of these three pixels.}
\end{figure}

\section{Discussion}
\label{sec:discussion} 
In this work we demonstrate the suitability of LASSO modified MLR as implemented through the LAR algorithm for covariate assisted interpolation of a univariate response in a pedological context.
The computational efficiency of the LAR algorithm is such that it is feasible to explore 500 unique, 35 observation subsets of a design matrix composed of 800 potential covariate terms, whereas the application of exhaustive search variable selection to this task would not have been computationally feasible.
While LAR is often applied to the exploration of potential model spaces composed solely of linear main effects it may also be applied to the exploration of potential model spaces which include both polynomial terms for covariates and terms for the interactions of two or more covariates implemented through products of these terms.  
Efron et al. [2004] illustrate the exploration of such a model space in their simulation study which compares LAR, LAR-LASSO and Stagewise solution paths obtained from a potential model space comprised of linear main effects, interaction terms and quadratic terms.  
In such cases, the LAR algorithm is executed upon a design matrix that includes appropriately recentred and rescaled columns for polynomial terms and interaction terms.
In our case study we expand 63 covariates to 2205 potential covariate terms by considering polynomial terms for all covariates up to polynomial order 4 and all possible pairwise linear interaction terms.
Pre-filtering this full design matrix to enforce a MCCM between covariate pairs of 0.95 results in a design matrix comprised of 800 potential covariate terms.
The L$_1$ penalty in LASSO regression allows for exploration of design matrices that include such highly collinear pairs of covariates.
In contrast, it would be advisable to discard a great deal more of these covariates to reduce the degree of collinearity in the design matrices examined prior to conducting the variable selection with OLS based approaches such as information criteria based stepwise variable selection.
Our concern regarding discarding numerous members of correlated pairs of covariates prior to conducting the variable selection appears justified in our case study.
The VSEPE distributions arising from models fitted to design matrices filtered to enforce a MCCM between covariate pairs of 0.4 are more dispersed about zero than the VSEPE distributions arising from models fitted to design matrices filtered to enforce MCCM between covariate pairs of 0.95.
Furthermore, it is the model averaged predictions of the models selected from exploration of training sets constructed from this less stringently pre-filtered design matrix that have the greatest coefficient of determination.
\newline
\newline
In our analysis we assume that covariate response correlations do not vary across the study area and so adopt a non-spatial regression approach.
That is, we assume spatially stationary regression coefficients as the first stage of modelling the spatial distribution of \%SOC.  
Spatially non-stationary linear regression coefficients may have added little here if some of the covariates varied in a spatially correlated manner.  
If there is spatial non-stationarity in a correlation between a covariate and some component of the response, this variation could well have be captured in our models by the selection of a polynomial term for the covariate in question were it also varying spatially.
If this were the case, it would be difficult to show one of the these two interpretations to be more valid than the other in the absence of information beyond the data we have for the case study site.
Given our primary objective of spatial interpolation of the response, the mechanism by which this interpolation is achieved (spatially stationary coefficients of polynomial terms or spatially non-stationary coefficients of linear terms) is less important than it would be if we were attempting to identify the pedological and edaphic processes that produce the observed distribution of \%SOC.
\newline
\newline
Limitations of the analysis presented here include the interpolation of the covariates to the locations at which the response was observed being accomplished via separate models before the variable selection is performed.
Further limitations stem from these interpolations being accomplished in a manner contingent upon the assumption of isotropic spatial dependence in the mean deviations of the covariates being realigned.  
By realigning the covariates by means external to the variable selection processes we, in effect, assume that the values we supply to the variable selection process are observed without error at the response locations.
However, we know that there was both uncertainty associated with the collection of the covariates and uncertainty associated with the interpolation of the covariates to the locations at which the response was observed.
The hierarchical Bayesian models for spatially misaligned data outlined by Banerjee et al. [2004] would be an interesting extension in this regard if these models could be extended to accomplish the variable selection task we have encountered.
The advantage of such an approach would be a more realistic propagation of uncertainty, including the uncertainty associated with the spatial realignment of the data layers, through the model hierarchy to that associated with the final full cover areal predictions rather than the more limited cross validation based estimation of the uncertainty associated with areal prediction that we calculate here.  
If this were combined with a Bayesian LASSO, where the shrinkage parameter could be assigned a hyperprior and estimated as part of the model structure, the need for cross validation would no longer be as strong but the computational challenge would likely be substantial.
Other penalized likelihood methods such as adaptive LASSO \citep{Zou2006}, SCAD \citep{Fan2001} and MCP \citep{Zhang2010} could all form interesting comparisons to the LASSO modified MLR we have fitted with the LAR algorithm in this work.
Further interesting comparisons could be conducted with Bayesian LASSO \citep{Park2008}, model-averaged Bayesian CART \citep{Chipman1998}, random forests \citep{Breiman2001}, boosted regression trees \citep{Friedman2002} and model-averaged Bayesian treed regression \citep{Chipman2002} with Bayesian LASSO implemented in the terminal node MLRs.

\section{Acknowledgments}
This work was funded by the CRC for Spatial Information (CRCSI), established and supported under the Australian
Government Cooperative Research Centres Programme. 
One of the authors (BRF) wishes to acknowledge the receipt of a Postgraduate Scholarship from the CRCSI.
We thank Vincent Zoonekynd for the R code to calculate Poincar\'{e} segment paths.

\section{Appendix A: Study Site and Field Methodology}
\label{App:A}
\subsection{Study Site}
The study site was a 137ha area of land on the Sustainable, Manageable, Accessible, Rural Technology (SMART) Farm of the University of New England near Armidale, New South Wales (NSW), Australia.
The north-west corner of the SMART farm had coordinates \ang{30;22;59}S \ang{151;35;23}E and the south-east corner of the SMART farm had coordinates \ang{30;27;26}S \ang{151;39;52}E.
The study site was situated at the base of Mount Duval (1393m \citep{NationalParksandWildlifeService2003}) and formed a part of the Uralla Plutonic Suite/Mount Duval Adamellite (acid porphyritic, hornblende-biotite monzogranite) and was characterized by yellow and brown chromosols \citep{Isbell2002} upon the hills with alluvial soils and siliceous sand complexes distributed along the drainage routes. 
The study site typically received 790mm of annual, summer dominant rainfall \citep{Garraway2011}. 
The maximum elevation within the study site was 1120m and the elevation range across the study site was less than 110m.
The study site consisted of selectively cleared native pasture containing some remnant vegetation and regrowth and had historically been grazed by sheep and cattle.
The south-east corner of the study site was situated at grid reference 371434E, 6632499N MGA GDA 94 Zone 56.
Being used to grow pasture and receiving in excess of 450mm of annual rainfall, soils in this area fell within the class of agricultural soils deemed to have the highest potential of any agricultural soils in NSW for sequestration of atmospheric carbon as Soil Organic Carbon (SOC) \citep{Chan2008}.   
A hill-shaded plot of a digital elevation model cropped to the approximate boundaries of the study site and an aerial photograph of the study site have been presented as the panels of Figure 4.

\begin{figure}[h!]
\centering
\makebox{\includegraphics[height = 0.9\textheight]{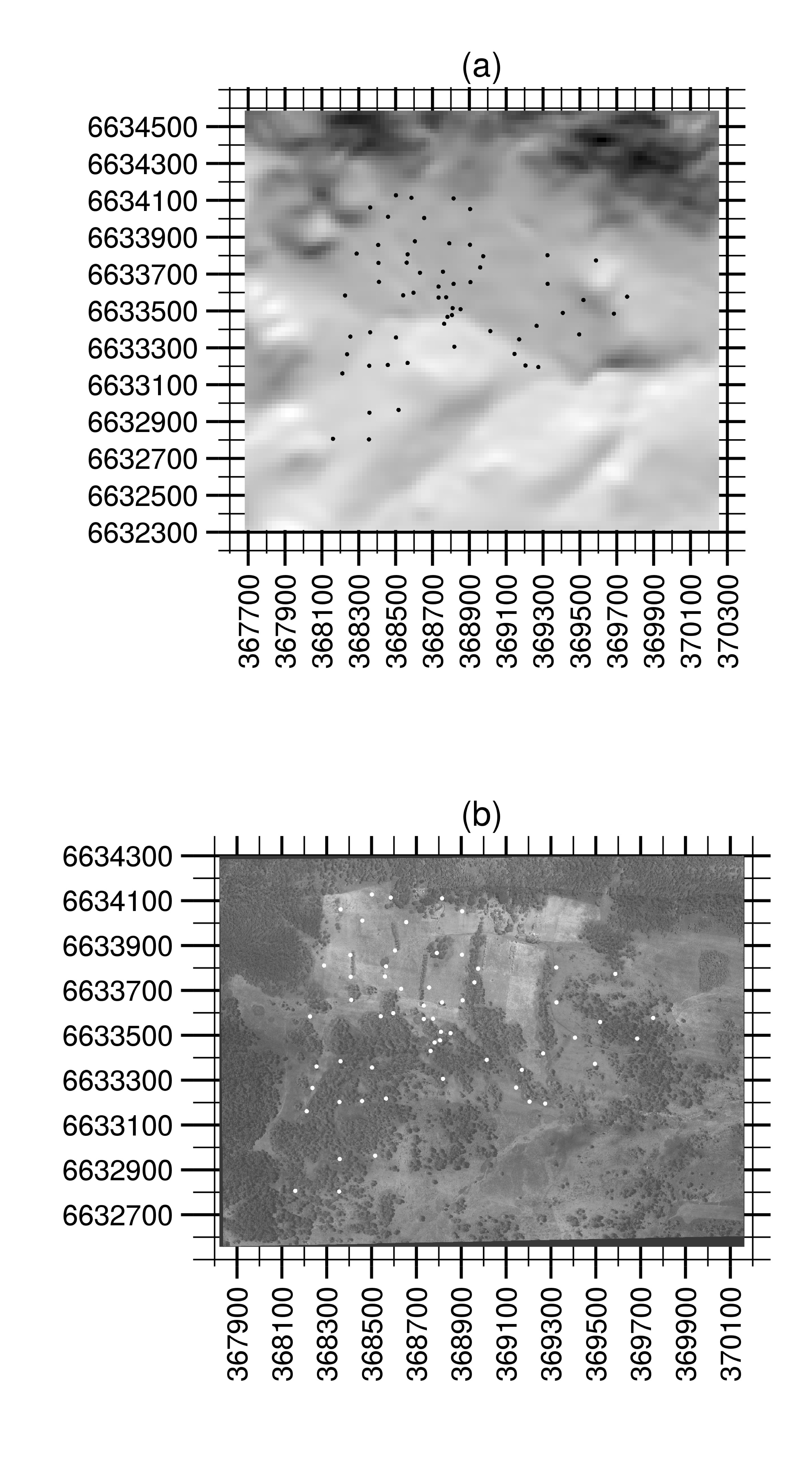}}
\caption{a) A hill shaded terrain surface for the study site calculated from a digital elevation model. b) An aerial photograph of the study site. The locations at which soil core samples were collected have been depicted as filled circles in both panels.}
\end{figure}

\subsection{Proximal Data Collection}
\subsubsection{Covariates}
The study area was surveyed three times in 2009 for soil apparent electrical conductivity ($EC_a$) and the reflectance of the top of the pasture canopy under active illumination.  
The first survey was conducted in the warm summer month of February, following a prolonged dry period (0mm of precipitation in the preceding seven days) in what was otherwise the second wettest month of the year.
The second survey was conducted in May, immediately after a significant rainfall event (84mm in the preceding seven days) in what was otherwise the cooler middle of the year when less rain fell than in summer.
The third survey was conducted in November, which marked the end of the winter growing season and was a month in which less rain fell than did in the wet month of October and the very wet month of December.  
These surveys were conducted by an all terrain vehicle (ATV) towing a sensory array that consisted of a specially configured Geonics EM38 unit (Geonics Ontario Canada), an LED illumination array, near-infrared and visible light reflectance sensors (Crop Circle$^{TM}$, Holland Scientific, USA) and a differential global positioning system (dGPS) (Trimble, Sunnyvale CA USA).  
\newline
\newline
The Geonics EM38 instrument measured soil $EC_a$ which may be understood as the integral of the electrical conductivity response recorded across soil depths.
To collect these data the EM38 instrument emitted a varying magnetic field which induced an electric current in the soil underneath the instrument.
The electric current induced in the soil created a magnetic field, the strength of which was proportional to the amount of electric current induced.
The strength of the magnetic field that resulted from the current induced in the soil was taken as indicative of the strength of this induced electric current and recorded by the instrument.
The strength of the electric current induced in the soil by the instrument-emitted electric field varied as a function of depth.
With a Geonics EM38 instrument operated in the vertical dipole orientation, as Garraway et al. orientated the instrument in their survey, the relative signal response, $S_v$, varied with depth, $z$, as follows $S_v(z) = \frac{4z}{(4z^2+1)^{\frac{3}{2}}}$ \citep{Morris2009}.
Due to air essentially not conducting electricity at these strengths of inductive magnetic fields, the conductivity signal commenced at the soil surface.
Thus the $EC_a$ measured by the EM38 instrument in vertical dipole mode was $\int_{H}^{\infty}\sigma(z)S_v(z)dz$ where $\sigma(z)$ was the electrical conductivity of the soil at depth $z$ and $S_v(z)$ was the relative signal response of the soil at depth $z$ \citep{Morris2009}.
Thus, when operated in vertical dipole mode, the Geonics EM38 instrument had a peak relative signal response to the electrical conductivity of the soil 350mm below the instrument and was essential insensitive to the electrical conductivity of any medium immediately below the instrument. 
Garraway et al. mounted a Geonics EM38 instrument on a rubber sled that held the instrument approximately 15mm above the ground \citep{Lamb2005} and towed this sled around the study area.
Thus the $EC_a$ data we have for the case study would have been dominated by the soil electrical conductivity at depths around the relative signal response peak at 335mm below the soil surface with very little contribution from the electrical conductivity of the soil surface.
\newline
\newline
The reflectance sensors measured top of pasture reflectance of active illumination in the Near InfraRed (NIR) and visible Red (RED) regions of the electromagnetic spectrum.
This style of proximal sensing of the reflectance of active illumination has been applied to both crop and soil mapping \citep{Holland2012}.  
The electro-optical principles governing the effectiveness of such sensors have been discussed in Holland et al. [2012].
The $EC_a$, RED reflectance and NIR reflectance were recorded simultaneously at regular time intervals as the ATV traversed the study area.
The east-west and north-south coordinates that accompanied each of these covariate observations were also recorded along with the associated Position Dilution Of Precision (PDOP) and Horizontal Dilution of Precision (HDOP) values.
The data from each of the ATV surveys were cleaned of observations with large inaccuracies in positioning (as assessed by HDOP and PDOP measurements).
The number of point observations that remained from each ATV survey after this cleaning had been conducted have been included in Table \ref{DataSummary}. 

\begin{table}
\caption{A summary of the types of data used in the case study, including for each type of data the: date of collection/publication, type of variable and number of observations or resolution (pixel dimensions).}
\label{DataSummary} 
\begin{tabular}{l l l l}
Variable            & Collected or      & Type           & Observations or \\
                    & Published         &                & Resolution \\
\hline 
\hline
SOC                 & 2009              & Geostatistical & 60     \\
\hline                       
$EC_a$ \& VI        & Feb 2009          & Geostatistical & 16179   \\
                    & May 2009          & Geostatistical & 16094   \\
                    & Nov 2009          & Geostatistical & 14059   \\
\hline
FPC                 & 2011 \& 2012      & Raster         & 10m$^2$ \\
\hline
DEM \& DEM Products & 2004              & Raster         & 25m$^2$ \\
\hline                       
Radiometric \&      & 2002 \& 2003      & Raster         & 50m$^2$ \\
Electromagnetic     &                   &                &  \\
Survey              &                   &                &  \\
\hline
\hline
\multicolumn{4}{ l }{ abbreviations: VI = Vegetation Indices, DEM = Digital Elevation Model, } \\
\multicolumn{4}{ l }{ FPC = Foliar Projective Cover, Feb = February and Nov = November.} \\
\end{tabular}
\end{table}

\subsubsection{Response}
In 2009, the study area was divided into five strata by means of k-means clustering \citep{Bishop2006} the red, green and blue channels from aerial imagery and the $EC_a$ data from the February survey \citep{Garraway2011}.
At least six locations for soil core sampling were randomly selected within each strata with additional locations manually selected to improve the representation of landscape attributes.
This stratified random sampling approach to choosing locations for soil core samples was similar to the process outlined in \citep{Miklos2010}.
As a result, soil samples were collected to a depth of 200mm at 60 locations across the study area with locations georeferenced using a differential Global Positioning System (dGPS) instrument.  
At each of the 60 dGPS coordinates, three soil cores were collected from within a 1m radius of the coordinates and aggregated to form a single soil sample that was laboratory analyzed for percentage SOC (hereafter \%SOC).  
Garraway et al. detailed the preparation of soil samples for assessment of the total organic carbon with a Carlo Erba NA 1500 solid sample analyzer (Carlo Erba Instruments, Milan, Italy).

\subsection{Remotely Sensed Data}
A 25m$^2$ resolution Digital Elevation Model (DEM) (sourced from the Department of Lands, New South Wales State Government, Australia) for the Armidale-Dumaresq region which contained the catchment in which the study area was situated was read into the System for Automated Geoscientific Analyses (SAGA v2.1.0 \citep{Conrad2015}) software to calculate the terrain topographic and hydrological attributes listed in Table \ref{CovCount}.
The ready availability of the attributes listed in Table \ref{CovCount} and potential relevance of topography and hydrology to SOC distributions generally, lead us to include the full suite of such attributes that may be calculated with SAGA as potential covariates in this analysis.  
The resulting GIS layers were then read into R \citep{R2015} with the `RSAGA' \citep{RSAGA2015} package.
Full cover layers for Foliar Projective Cover (FPC) produced by applying the Statewide Landcover and Trees Study (\href{https://www.qld.gov.au/environment/land/vegetation/mapping/slats-methodology/}{https://www.qld.gov.au/environment/land/vegetation/mapping/slats-methodology/}) method to imagery from the SPOT5 satellite (10m$^2$ resolution) were acquired for the study region from 2012 and 2011 from the New South Wales State Government Department of Environment.
These layers were read into R \citep{R2015} with the `raster' package \citep{Hijmans2012}.
Potassium (K), Uranium (U) and Thorium (Th) count layers from an airborne $\gamma$ ray radiometric survey \citep{Brown2002} and similar layers from six channels of electromagnetic imagery \citep{Brown2003} were read into R\citep{R2015} with the `raster'\citep{Hijmans2012} package.
\clearpage
\begin{table}
\caption{The 63 potential covariates from which models were built for soil organic carbon in this study.}
\label{CovCount} 
\begin{tabular}{ p{3.5cm} p{7cm} p{2cm} }
Source & Covariate Name & Acronym \\
\hline
\hline
ATV Top of Pasture & Soil Apparent Electrical Conductivity & ECA \\
Surveys & Near InfraRed Reflectance & NIR \\
12 covariates & Red Reflectance & RED \\
from each of February, & Simple Ratio & SR \\
May \& November & Difference Vegetation Index & DVI \\
 = 36 covariates & Normalized Difference Vegetation Index & NDVI \\
 & Soil Adjusted Vegetation Index & SAVI \\
 & Non-Linear Vegetation Index & NLVI \\
 & Modified Non-Linear Vegetation Index & MNLVI \\
 & Modified Simple Ratio & MSR \\
 & Transformed Vegetation Index & TVI \\
 & Re-normalised Difference Vegetation Index & RDVI \\
\hline
Terrain \&& Catchment Area & CatAr \\
Hyrdology Metrics & Catchment Height & CatHe \\
Calculated from a & Catchment Slope & CatSl\\
$25m^2$ resolution & Cosine(Aspect) & CosAsp \\
DEM & Elevation &  Elev \\
= 16 Covariates & Slope Length Factor & LSF \\
& Plan Curvature & PlanC \\
& Profile Curvature & ProfC \\
& Sky View Factor & SVF \\
& Slope & Slp \\
& Stream Power Index & SPI \\
& Terrain Ruggedness Index & TRI \\
& Topographic Position Index & TPI \\
& Vector Terrain Ruggedness & VTR \\
& Visible Sky & VS \\
& Wetness Index & WI \\
\hline
Foliar Projective Cover & 2011 & FPCI\\
Layers = 2 Covariates & 2012 & FPCII \\
\hline
Electromagnetic & 1 & MagI \\
Imagery Channels & 2 & MagII \\
= 6 Covariates & 3 & MagIII  \\
& 4 & MagIV  \\
& 5 & MagV \\
& 6 & MagVI \\
\hline
$\gamma$ Radiometric Layers & Potassium & K \\
= 3 Covariates & Thorium & Th \\
 & Uranium & U \\
\hline
\hline
\end{tabular}
\end{table}
\clearpage
Cartographic projection systems are used to project latitude and longitude coordinates from a particular region of the surface of the Earth onto a two dimensional plane.  
In this analysis all spatial coordinates were treated as coordinates on a two dimensional plane and as such it was important to ensure that all data layers utilised the same projection system.  
This was accomplished with the R\citep{R2015} package `raster'\citep{Hijmans2012} through the re-projection of all data layers that did not already use the most common projection system among the data layers to this projection system, namely a UTM projection for zone 56 South using the WGS84 ellipse.

\section{Appendix B: Choice of Covariates}
\label{App:B}
In this Appendix we explain our choice of environmental characteristics considered as potential covariates for modelling soil carbon.
The first stage in collating this set of potential covariates was to identify the common covariates used in soil carbon modelling via a review of the available literature.
Once we had this list of potential covariates, the second stage was to identify which of these we could obtain for our case study site.

\subsection{Soil Organic Carbon and Soil Apparent Electrical Conductivity} 
\label{sec:potential.covariates}
Soil apparent electrical conductivity ($EC_a$) has variously been found indicative of some or all of soil: moisture content, pore distribution, pore size, salinity, clay content, mineralogy, cation exchange capacity and temperature \citep{Barnes2003}.  
Discretizing soil $EC_a$ values into four classes produced significant factors in separate ANOVAs for each of soil total organic matter, soil particulate organic matter and total soil carbon in the top 30cm of soil sampled across 250 ha of farmland used for wheat, corn and millet in the state of Colorado in the United States of America (USA) \citep{Johnson2001}. 
A negative correlation was detected between soil $EC_a$ and SOC (r = -0.42) in the top 90cm of soil in 9ha of farmland that had a long history of cotton row cropping in Alabama, USA but no such correlation was detected in the top 30cm \citep{Terra2006}.  
Soil $EC_a$ was more correlated with soil carbon (r = -0.65 to -0.76) than were any of the local relative elevation, local slope and satellite measured soil surface reflectance of near bare fields at three sites (48.7 ha, 52.4 ha and 65.4 ha in area respectively) in farmland used for maize and soybean cropping in Nebraska, USA \citep{Simbahan2006}.
Studies have also been performed where it seemed likely that influences other than SOC were dominating the soil $EC_a$ signal.  
Principal Component Analysis (PCA) of soil properties in an 8ha agricultural field in Flanders, Belgium yielded largely independent spatial patterns in soil pH, $EC_a$ and SOC \citep{Vitharana2008}.  
Furthermore, in the soils of the world's oldest continuous cotton experiment (Alabama, USA) little correlation was detected between SOC in the top 15cm of soil and the $EC_a$ of the top 30cm of soil across 0.4ha of land \citep{SiriPrieto2006}.

\subsection{Soil Organic Carbon and Spectral Vegetation Indices} 
\label{App.A:SOC_VI}
Land plant biomass in any location will have been influenced by many soil conditions.
Through direct and indirect effects on soil: structural stability, water and nutrient retention, faunal activity and diversity, and elemental recycling \citep{Lal2009} SOC levels may have influenced land plant biomass in many situations.
Conversely, plants will have also provided an input of carbon to SOC via litter fall and root turn over.
Thus empirical correlations between \%SOC and plant biomass are plausible.
The amount of green plant biomass present in a location may be indicated by the density of green leaves present above the soil there.
The density of green leaves present above the land surface has often been estimated from the reflection spectra of the land surface when observed from above canopy height.  
Three spectral signatures of particular interest for these considerations have been identified as: those of healthy green leaves, those of stressed or senescent green leaves and those of agricultural soils. 
The marked difference in the intensities of light reflected from green leaves in the visible red (RED) and near infra red (NIR) wavelengths and the general weakness or absence of such a `red edge' in the spectral reflectance signatures of stressed or senescent leaves and agricultural soils has formed the basis of many spectral vegetation indices used for monitoring vegetation \citep{Pinter2003}.  
Healthy leaves have typically exhibited a high reflectance of NIR light due to scattering of these wavelengths at the interface between the mesophyll and cell walls and low absorption of these wavelengths by photosynthetic pigments and organelles \citep{Pinter2003}.  
Whereas, healthy green leaves have typically displayed a low reflectance of visible wavelengths due to the high absorption of light in this region of the spectrum by photosynthetic and accessory pigments \citep{Pinter2003}.  
Green plant stress and senescence have often manifested in the form of depressed chlorophyll concentrations and the expression of accessory leaf pigments which together lower the absorption of visible wavelengths by leaves. 
Where stress or senescence has lowered the absorption of visible wavelengths by leaves the reflectance peak of such leaves will have widened correspondingly from the green region of the spectrum typical of healthy green leaves through towards redder wavelengths.  
Where stress or senescence has manifested in this broadened reflectance of visible spectra by leaves a simultaneous decrease in the NIR reflectance of these leaves will have also occurred.
Thus where it has occurred, stress or senescence would have resulted in a loss or weakening of the abrupt `red edge' typical of the reflectance spectra of healthy green leaves \citep{Pinter2003}.
Similarly, agricultural soils have been characterised by a lack of any such sharp contrast in the reflectance intensities of different wavelengths \citep{Pinter2003}. 
Most spectral vegetation indices have been constructed as functions of NIR and RED reflectance designed to quantify some aspect of the expected differences between the reflectance spectra of healthy green leaves and those of soils and stressed or senescent leaves.
Thus vegetation indices have been designed to provide an estimate of the quantity of green plant material that contributed to the reflectance spectra of a land surface by quantifying the extent to which such differences in NIR and RED intensities occurred within this spectra \citep{Pinter2003}.  
The spectral signature obtained from an entire canopy may have differed markedly from that obtained from an individual green leaf and furthermore may have varied across the growing season as the canopy geometry altered with plant growth \citep{Pinter2003}.
Thus differences between vegetation indices calculated from reflectance surveys conducted at the same location at different stages in the year may have yielded information about how green plant biomass there changed across the growing season.
Changes in biomass across a growing season could, in turn, have been indicative of soil attributes baring other stronger influences on plant biomass lost or accrued.
Thus the comparison of vegetation index values when collected in the same location at different points in the plant growth cycle could have been indicative of soil properties.
For instance, plants that grew in poorer soils and experienced otherwise similar conditions could reasonably be expected to have produced less biomass in a growing season than the same plants in more favourable soils. 
Furthermore, the presence of SOC in surface soils has been documented as increasing soil aggregation thereby creating larger lacunae (also referred to as interstitial spaces) into which water may drain from the soil surface.
Thus SOC has been broadly classified as beneficial to the infiltration of soil by water and the retention of water by soil\citep{Franzluebbers2002}.  
Thus increased \%SOC at a location could in turn aid water retention there and allow plants that undergo seasonal curing (e.g. grasses like those at our case study site) to remain green longer into the dry part of the year at that location.
\newline
\newline
A review of studies of the correlations between soil organic matter and crop reflectance in visible and NIR regions of the electromagnetic spectrum and the vegetation indices thereby derived formed a section of the review paper \citep{Barnes2003}. 
In certain situations, above ground plant biomass may have been related to SOC concentrations.
Where this has been the case vegetation indices may have held relevance to \%SOC concentrations.
This seems to have been the case in the following studies.
The Normalised Difference Vegetation Index (NDVI \citep{Rouse1973}) and the Soil Adjusted Vegetation Index (SAVI \citep{Huete1988}) have both achieved considerable popularity as vegetation indices.
A positive correlation between canopy NDVI and biomass was detected in a 7ha cotton field in Larissa, Greece \citep{Stamatiadis2005}.  
Furthermore, the pasture canopy SAVI was found to be the best of a range of vegetation spectral indices for predicting pasture green dry mass across four 50ha paddocks in New South Wales, Australia \citep{Trotter2010}.  
Much like soil $EC_a$, plant visible and NIR reflectance have variously been found correlated with soil properties other than SOC such as soil moisture and Cation Exchange Capacity (CEC) \citep{Barnes2003} along with prevailing climate, ecosystem, terrain and physical soil properties \citep{Mulder2011}.
We have summarised studies that found correlations between any of a selection of vegetation indices that may be calculated from reflectance intensities in the NIR and RED bands and soil carbon in Table \ref{VISOC}.
Since our study site consisted of native pasture with remnant woody vegetation we restricted this summary to one of studies that were conducted in pastures, grasslands, prairies and steppes.

\begin{table}
\caption{Vegetation Indices reported to have been correlated with grass biomass (which were thus possibly also correlated with soil carbon) in pastures, grasslands, prairies and steppes.}
\label{VISOC}
\begin{tabular}{ p{1cm} p{4cm} p{3.25cm} p{2cm} }
VI & Full Name & Formula & Correlation with Grass Biomass \\
\hline
\hline
 &  &  &  \\
SR     & Simple Ratio & $ \frac{NIR}{RED} $ & \citep{Trotter2010} \\
 &  &  &  \\
DVI    & Difference Vegetation Index & $ NIR - RED $ & \citep{Munoz-Robles2012, Payero2004} \\
 &  &  &  \\
NDVI   & Normalized Difference Vegetation Index & $ \frac{NIR - RED}{NIR + RED}$ & \citep{Trotter2010, Munoz-Robles2012} \\
 &  &  &  \\
SAVI   & Soil Adjusted Vegetation Index* & $ \frac{(NIR - RED)(1+L)}{NIR+RED+L} $ & \citep{Trotter2010} \\
 &  &  &  \\
NLI    & Non-Linear Vegetation Index & $ \frac{NIR^2 - RED}{NIR^2 + RED} $ & \citep{Trotter2010} \\
 &  &  &  \\
MNLI   & Modified Non-Linear Vegetation Index & $ \frac{(NIR^2 - RED)(1+L)}{NIR^2+RED+L}  $ & \citep{Trotter2010} \\
 &  &  &  \\
MSR    & Modified Simple Ratio & $ \frac{\frac{NIR}{RED} - 1}{(\frac{NIR}{RED})^{1/2} +1} $ & \citep{Trotter2010} \\
 &  &  &  \\
TVI    & Transformed Vegetation Index & $ (NDVI + 0.5)^{1/2} $ & \citep{Payero2004} \\
 &  &  &  \\
RDVI  & Re-Normalised Difference Vegetation Index & $ \frac{NIR - VIS}{(NIR + VIS)^{1/2}} $ & \citep{Payero2004} \\
 &  &  &  \\
\hline
\hline
\multicolumn{4}{ l }{* L = 0.5 recommended for wide range of leaf area index values}
\end{tabular}
\end{table}

\subsection{Soil Organic Carbon and Scattered Paddock Trees} 
Below ground root turnover and above ground litter fall have both been recognised as sources of detrital carbon to topsoil.
Thus in native pastures with remnant woody vegetation in the form of scattered paddock trees, such as the pastures from which our case study data were collected, the locations of these trees may have influenced the spatial distribution of SOC.
Elevated concentrations of organic matter in soils beneath and around trees and shrubs relative to surrounding soils have been observed across a variety of environments and ecosystems\citep{Hibbard2011,Graham2004}.
In the Northern Tablelands of New South Wales (the region from which the data analysed here were collected) an ANOVA detected significantly elevated ($P < 0.001$) organic carbon content in the top 5cm of soils underneath the canopies of scattered paddock trees compared to soils beyond these canopy margins \citep{Graham2004}.

\subsection{Soil Organic Carbon and Digital Elevation Model Derived Terrain Descriptors}  
Climate, parent material, topography and biotic factors may all have influenced pedogenesis to varying degrees in different ecological and geographic contexts.  
Topography may have influenced soil characteristics to a greater or lesser extent by having influenced hydrologic and erosional processes (e.g. soil water content, runoff and sedimentation) along with soil temperature (via aspect, exposure etc.) which together form and alter soils through mineral weathering, erosion, leaching, decomposition, horizontal zonation and sedimentation \citep{Moore1993}. 
Topography may also have affected the process of SOC loss that accompanied the conversion of natural land into agricultural land by having influenced SOC: leaching, movement as dissolved organic carbon or particulate organic carbon suspended in water flowing over or through the soil, and erosion by wind or water runoff moving soil and the constituent SOC \citep{Lal2002}. 
For the purposes of geostatistical modelling, topography has often been quantified via terrain metrics (e.g. elevation, slope, aspect, curvature, etc.) and hydrological metrics (e.g. catchment area, soil wetness index, stream force index, etc.) calculated for each of the pixels in a digital elevation model (DEM) of the land surface.
\newline
\newline
In 460ha of cropping and pastoral land in north-west New South Wales, Australia regression modelling identified elevation and plan curvature along with $EC_a$, $\gamma -$ray radiometric potassium and thorium related emissions as useful predictors of total soil carbon \citep{Miklos2010}.  
In a 9ha field with a long history of row crop monoculture subject to conventional tillage in central Alabama, USA SOC was found correlated with a compound topographic index (metric of potential for water pooling on the land surface) (r = 0.48) and with land slope (r = -0.42) which lead the authors to postulate that erosion and field scale hydrodynamics were likely responsible for a large portion of the variability detected there in soil carbon \citep{Terra2005}.  
In mapping soil carbon in a 12.5ha field with a history of crop rotation between corn and soy beans in central Michigan, USA models that utilised terrain slope, aspect, plan curvature, profile curvature and tangential curvature were generally found to perform better than those that did not \citep{Mueller2003}. 
In 9ha of cropping soil typically used for cotton in Alabama, USA models with combined topographic index, elevation, slope, silt content and $EC_a$ as covariates were found to account for up to 50\% of the SOC variability leading the authors to conclude that the spatial distribution of SOC had been affected prominently by topography and historical erosion \citep{Terra2004}.  
In this same farmland in 2006 SOC concentrations in the top 30cm of soil were found to be correlated with the composite terrain index (r = 0.48) and terrain slope (r = -0.41) \citep{Terra2006}.  
In a 4.2ha catchment used as agricultural land in North Rhine-Westphalia, Germany correlations between SOC and profile curvature, plan curvature, catchment area, stream power index \citep{Moore1993} and predictions from water and tillage erosion models \citep{Quinn1991, VanOost2000, VanRompaey2001} of soil redistribution patterns have been detected \citep{Dlugoss2010}.  
From a study of 5.4ha of a dryland agroecosystem with a long history of winter wheat in North-Eastern Colorado, slope and wetness index were identified as the terrain attributes most correlated with soil organic matter \citep{Moore1993}.  
Variable selection in this same study returned a linear model that explained 48\% of soil organic matter variation with the covariates wetness index, stream power index and aspect.
Studies that detected correlations between soil carbon and a selection of topography and hydrology metrics that may be calculated with the SAGA \citep{Conrad2015} have been summarised in Table \ref{SAGASOC}.

\begin{table}[h!]
\caption{Topographic and hydrological metrics reported to have been correlated with soil carbon in agricultural land.}
\label{SAGASOC} 
\begin{tabular}{ p{4cm} p{4cm} }
Metric & Correlation with \\
 & Soil Carbon \\
\hline
\hline
Aspect & \citep{Moore1993, Mueller2003} \\
& \\
Catchment Area & \citep{Dlugoss2010} \\
 & \\
Elevation & \citep{Terra2004, Miklos2010} \\
 & \\
Plan Curvature &  \citep{Moore1993, Mueller2003, Dlugoss2010, Miklos2010} \\
 & \\
Profile Curvature & \citep{Moore1993, Mueller2003, Dlugoss2010} \\
 & \\
Slope & \citep{Moore1993, Mueller2003, Terra2004, Terra2005, Terra2006} \\
& \\
Stream Power Index & \citep{Moore1993, Dlugoss2010} \\
 & \\
Tangential Curvature & \citep{Mueller2003} \\
 & \\
Topographic Indices &  \citep{Terra2004, Terra2005, Terra2006} \\
 & \\
Wetness Index & \citep{Moore1993} \\
\hline
\hline
\end{tabular}
\end{table}

\subsection{Soil Organic Carbon and Radiometric Imaging of the Earth Surface}
Recording $\gamma$ radiation naturally emitted from the surface of the Earth has been established as a means to detect geochemical anomalies in particular those associated with ore bodies \citep{Cook1996}.
Collecting aerial images of such $\gamma$ radiation emissions has been termed $\gamma$ radiometry and the spectral signatures most frequently observed have been those associated with the production of $^{238}U$, $^{232}Th$ and $^{40}K$ daughter radionuclides \citep{Cook1996}.  
In addition to detecting minerals rich in Uranium and Thorium and mapping geology based on prior knowledge of associations between the above radionuclides and geological materials, $\gamma$ radiometry of a landscape has also facilitated the tracking of geochemical anomalies and inference regarding erosional processes therein \citep{Cook1996}.
Such links to pedological processes have enabled $\gamma$ radiometry to be used for soil mapping \citep{Cook1996}.
On a broad spatial scale airborne radiometric data (particularly the K band) was found to improve the mean square error of predictions of soil organic carbon across Northern Ireland ($\sim$13,843 km$^2$) when coupled with elevation data to 30.6\% \citep{Rawlins2009}.  
Similarly, digital elevation model derived soil properties and $\gamma$ radiometric survey data when used to build regression trees were found to account for 54\% of the total soil carbon variation across 50, 000 ha of state forest in south eastern Australia \citep{McKenzie1999}.
Furthermore, $\gamma$ radiometry has also been found useful for predicting the spatial distribution of soil carbon on scales closer to that across which our data were collected. 
Over a 5625ha square area of cropping land on the lower plains of the Macquarie River, News South Wales (Australia) percentage soil organic carbon in the top soil was found to be weakly negatively correlated with the concentrations of Potassium and Uranium in the ground as measured by a $\gamma$ radiometric survey \citep{Singh2013}.
Furthermore, regression modelling identified elevation and plan curvature along with $EC_a$, $\gamma -$ray radiometric potassium and thorium reflectances as useful predictors of total soil carbon in 460ha of cropping and pastoral land in New South Wales \citep{Miklos2010}.

\section{Appendix C: Choice of Modelling Method}
\label{App:C}
In this appendix we compare and contrast a selection of Multiple Linear Regression (hereafter MLR) and Binary Tree (hereafter BT) based techniques in the context of data and objectives akin to those of our case study. 
We consider the defining characteristics of our case study data to be: (1) more potential covariate terms than observations (the $p > n$ or ultrahigh dimensional situation for variable selection)  (2) a high degree of collinearity among the potential covariate terms and (3) suspected importance of non-linear effects of covariates and interactions of covariate effects.
The primary objective of our case study analysis was covariate assisted spatial interpolation of the response.
Our case study also had the additional context of the modest computational resources provided by one mid-range laptop and our desire for an easily interpretable predictive mechanism.
In Section \ref{SS:BT} we introduce a selection of BT based models and in Section \ref{SS:MLR} we introduce a selection of MLR based models.
In Section \ref{SS:MLRcfBT} we compare the relative merits of the model introduced in Section \ref{SS:BT} and Section \ref{SS:MLR} for modelling data like those of our case study with our objective of covariate assisted spatial interpolation of the response.

\subsection{An Introduction to the Binary Tree Based Models Considered in this Work}
\label{SS:BT}
\subsubsection{CART}
Classification And Regression Trees (CART) \citep{Breiman1984} are two closely related techniques.
Both utilise a binary tree based model structure but classification trees predict a categorical response while regression trees predict a continuous response.
Our interest here is in techniques for modeling a continuous response and thus methods for solving regression problems.
Subsequently we focus on regression trees.
The regression trees of Breiman et al. 1984 partition the response observations $y_i$, $i = 1, ..., n$, into $M$ mutually exclusive sets by recursively partitioning the associated covariate space into $M$ mutually exclusive regions $R_1, ..., R_M$ through binary divisions along covariate axes.
Each associated subset of the response is then modelled by the single parameter, maximum likelihood estimator for those observations, the group mean $\hat c_m$ \citep{Hastie2009}.
Thus, regression trees model a continuous response as per Equation \ref{Eq:RT}.
\begin{equation}
\label{Eq:RT}
\hat y_i = \sum_{m = 1}^M \hat c_m I(\mathbf{x}_{i,.} \in R_m)
\end{equation}
Each $y_i$ will be modelled by one and only one $\hat c_m$ since the $R_m$ are disjoint and thus for any particular $i$, $\mathbf{x}_{i,.} \in R_m$ for exactly one unique $R_m$.
\newline
\newline
Thus when continuous covariates are supplied to regression trees the predictions produced vary in a stepwise manner across the range of the covariates whereas the predictions from an MLR supplied with these same covariates would vary in a continuous manner across this same range.
Furthermore, the recursive nature of the binary partitions of regression trees enable far more complex interactions to be modelled than those permitted by taking products of pairs of covariates as may be done in MLR.
However, deep trees (trees with many binary partitions) would be required to form good approximations to even simple linear relationships between covariates and the response whereas such relationships may be modelled naturally as components of the structure of MLR.
Deep trees rapidly become a concern with a limited number of response observations since they result in terminal node parameters being estimated from fewer observations (and thus less reliably).

\paragraph{Frequentist CART}
Regression trees, as proposed by Breiman et al. 1984, are constructed via recursive binary partitions of the observations based on whether particular covariate values of these observations exceeded threshold values with the range of these covariates.
This results in the $M$ mutually exclusive regions of covariate space $R_1, ..., R_M$ referred to in Equation \ref{Eq:RT}.
Within each particular $R_m$ the $\hat c_m$ that minimises the residual sum of squares for predicting the response in that region is simply the group mean for those response observations $\hat c_m = E(y_i|x_i \in R_m)$.
The challenge when fitting regression trees is identifying a sequence of binary partitions that define a set of regions such that the residual sum of squares from the entire tree is low.
Since an exhaustive search of the potential regression trees that may be constructed from a particular set of data is computationally infeasible in all but the most trivial cases, regression trees are typically constructed via a greedy algorithm in the hope of identifying a regression tree that fits the data well via a computationally tractable procedure \citep{Hastie2009}.
Greedy algorithms are so named because at each step in the iterations of such an algorithm the decision that yields the best improvement in the decision metric (e.g. fit of the model etc.) between the current state and the next is the decisions that is taken.
As such there is no long term planning or stochasticity involved in such algorithms and thus no guarantee of identifying the optimal solution.
Indeed, such algorithms can be seen to be highly sensitive to local maxima.
\newline
\newline
For fitting regression trees using the residual sum of squares as the criterion for decisions to partition the data, the greedy algorithm is as follows.
The algorithm commences with all data in a single set termed the root node of the tree.
All possible binary divisions of this root node set along all possible covariate axes are then constructed in turn, and the residual sums of squares resulting from prediction of the response with the pairs of associated group means are computed.
With a finite number of observations there is a finite number of ways to divide the data into two subsets based on the covariate values of these observations.
This complete but finite set of possible divisions of the data may be obtained by considering threshold values equidistant between each of the pairs of observed values of a covariate when these values are arranged in an ascending (or descending sequence) for each covariate in turn.
The binary partition that yields the best improvement in the residual sum of squares for the entire tree relative to that at the previous step is then selected as the partition to use at this step in the algorithm and the process is repeated for each of the resulting subsets (also referred to as child nodes).
This recursive partitioning process is then continued until some stopping criterion is met.
A simple and popular choice of stopping criterion is a minimum number of observations per terminal node from which the practitioner considers it is still reliable to estimate a mean.
Once this tree growing algorithm is halted by the satisfaction of the selected stopping criterion the resulting tree may then be `pruned' by sequentially examining the effects of collapsing the parent nodes of the current terminal nodes on the basis of some \textit{cost-complexity} criterion and taking this action where it is judged meritorious.
More formally this growing and subsequent pruning procedure may be described as follows.
\newline
\newline
The algorithm commences at the root node which contains all observations.
Given the full set of covariates as the design matrix, $X$, the algorithm takes each covariate, $X_{j}$, in turn and computes the set of threshold values that would each produce different subsets of the response should they be used to define a binary partition of the data based on this covariate.
For each unique pairing of a particular covariate, $X_{j}$, and a particular threshold value for that covariate, $s$, the regions defined by a binary partition based on $X_j$ and $s$ are two disjoint regions of covariate space:
$R_1(j,s) = \{ X|X_j \leq s \}$ and $R_2(j,s) = \{ X|X_j > s \}$.
The algorithm compares all such regions that may be constructed at this step to identify the choice of covariate $X_{j}$ and threshold value $s$ that solves
\begin{equation} 
\argmin\limits_{j,s} \Big( \argmin\limits_{c_1} \big( \sum\limits_{x_{i,j} \in R_1(j,s)} (y_i - \hat c_1)^2 \big) + \argmin\limits_{c_2} \big( \sum\limits_{x_{i,j} \in R_2(j,s)} (y_{i} - \hat c_2)^2 \big) \Big)
\end{equation}
where $\hat c_1$ and $\hat c_2$ are calculated for each pairs $(j,s)$ as the respective response group means:
$\hat c_1 = E \big( y_i | x_{i,j} \in R_1(j,s) \big)$ 
and 
$\hat c_2 = E \big( y_i | x_{i,j} \in R_2(j,s) \big)$ .
The resulting binary partition of the data is then made and the above process is repeated for each of the resulting child nodes until some stopping criterion, such as a threshold minimum number of observations per terminal node, is satisfied at which point the algorithm is halted.
The resulting tree may then be pruned recursively subject to the changes in some \textit{cost-complexity} criterion that result from collapsing the parent nodes of the various current terminal nodes.
More formally, let: $T \subset T_0$  be defined as any tree that may be obtained by collapsing some non-terminal node(s) of $T_0$, $|T|$ be defined as the number of terminal nodes in tree $T$ and let $m$ index the terminal nodes of $T$ each with the associated subset of the data $R_m$.
A \textit{cost-complexity} criterion, $C_\alpha(T)$, may be defined as per Equation 6.
\begin{align}
\label{Eq:PrunCCC}
N_m &= count\{x_i \in R_m\},\\
\hat c_m &= \frac{1}{N_m} \sum\limits_{x_i \in R_m} y_i,\\
Q_m(T) &= \frac{1}{N_m} \sum\limits_{x_i \in R_m} (y_i - \hat c_m)^2\\
C_\alpha(T) &= \sum\limits_{m=1}^{|T|} N_m Q_m(T) + \alpha |T|, \alpha \geq 0
\end{align}.
This \textit{cost-complexity} criterion is the sum of the residual sum of squares for the predictions from the whole tree and a multiple, $\alpha$, of the number of terminal nodes in the tree.
The tuning parameter $\alpha$ controls the trade off between the fit of the tree to the data and the complexity of the tree as quantified by the number of terminal nodes of the the tree.
Increases in $\alpha$ will yield smaller trees with larger residual sums of squares.
Each pair of original tree $T_0$, grown as above, and tuning parameter value $\alpha$ will have some smallest sub-tree $T_\alpha$ that minimizes $C_\alpha(T)$.
This $T_\alpha$ may be identified by \textit{weakest link pruning} \citep{Hastie2009} whereby the internal nodes that yield the smallest per-node increase in the residual sum of squares $ \sum\limits_{m=1}^{|T|} N_m Q_m(T) $ when collapsed are sequentially collapsed until there are no longer any binary partitions to collapse as the entire data are once again contained in the original `root' node associated set.
It has been shown that the sequence of sub-trees thus obtained dependably contains $T_\alpha$ \citep{Hastie2009,Breiman1984,Ripley1996}.
For the purposes of building a regression tree for interpolation, $\alpha$ may be estimated via cross validation \citep{Hastie2009}.
This is traditionally how regression trees have been fitted however there is also an approach extant for fitting regression trees under the Bayesian paradigm.

\paragraph{Bayesian CART}
The Bayesian approach to fitting CART models was developed by Chipman et al. (1998) and involves the use of particular prior specifications for the terminal node parameters and the tree structure itself along with a stochastic search.
The Bayesian CART \citep{Chipman1998} is a CART model, it is the manner in which the model is fitted and the underlying assumptions that are Bayesian.
Continuing our focus on models for a continuous response variable with continuous covariates we will describe the Bayesian approach to fitting a regression tree.
The Bayesian regression tree model consists of a binary tree $T$ with $b$ terminal nodes and the associated parameter vector $ \Theta = (\theta_1, \theta_2, ..., \theta_b) $, each $ \theta_i $ being associated with the $i$th terminal node of the tree.
If $\mathbf{x}_{i,.} = [ x_{i,1}, x_{i,2}, ...,x_{i,p} ]$ falls within the region defined by the $i$th terminal node, the associated response variable $y_i|\mathbf{x}_{i,.}$ is modeled by the distribution $f(y_i|\theta_i)$ with $f$ representing some parametric family controlled by parameter(s) $\theta_i$.
Chipman et al do not specify a closed form prior for the binary tree structure but instead specify it implicitly by generating trees from a tree growing stochastic process.
In this manner each tree grown by the stochastic process forms a randomly drawn observation from this tree prior.
Such specification of the tree prior allows for simple evaluation of prior probability $p(T)$ for any tree $T$ which in turn may be employed within a Metropolis Hasting (MH) algorithm.
To draw an observation from the tree prior a new tree is propagated from the tree consisting of a single `root' node by stochastically dividing terminal nodes in an iterative process.
This tree propagating process is governed by a function that controls the selection of terminal nodes for division and another function that controls the assignment of a division rule to a terminal node that has been selected to be divided.
The function $p_{DIVIDE}(\eta, T)$ generates the probability for tree $T$ that terminal node $\eta$ is divided.
If terminal node $\eta$ of tree $T$ is chosen to be divided, the function $p_{RULE}(\rho|\eta, T)$ generates the probability of assigning the division rule $\rho$ to this terminal node.
A division rule for a binary partition specifies the covariate that will be used to define the binary partition and the threshold value of this covariate which will determine the division of the observations into two groups based on the values of this covariate associated with each of the observations.
The form and parameter values assigned to these functions collectively control the frequency with which particular tree depths and geometries are generated and thus the eventual weighting of such trees in the prior distribution.
As such, via influence on the prior, these functions may be used to guide the posterior towards identifying trees of the desired depths and geometries (for instance to emphasise trees with a minimum number of observations in each terminal node).
The priors for the parameters of the distribution functions used to model the response observations in the terminal nodes may be taken as standard conjugate forms.
A convenient option for priors on the parameters $\Theta = (\theta_1, \theta_2, ..., \theta_b)$ for the terminal nodes $1, ..., b$ is to use mean shifted normal distributions for each $\theta_i$ with $(\mu_1, \mu_2, ..., \mu_b) | \sigma, T$ iid  $\sim N(\bar \mu, \frac{\sigma^2}{\alpha})$
and 
$\sigma^2 | T \sim IG(\frac{\nu}{2}, \frac{\nu \lambda }{2})$.
Such a formulation permits each terminal node an individual mean parameter and models all node mean parameters with independent and identical Normal distributions.
Should a more flexible formulation be desired an individual variance parameter may be introduced for each terminal node mean via mean-variance shifted normal distributions as follows i.e. for $i = 1, 2, ..., b$, $\mu_i | \sigma_i, T$ iid  $\sim N(\bar \mu, \frac{\sigma_i^2}{\alpha})$ and $\sigma_i^2 | T \sim IG(\frac{\nu}{2}, \frac{\nu \lambda }{2})$.
Both the above the priors $p(\Theta | T)$ facilitate closed from solutions for $p(Y|X,T) = \int p(Y|X,\Theta,T)p(\Theta|T)d\Theta$ to be obtained analytically.
Utilization of one of these closed form solutions along with a CART tree prior $P(T)$ enables the posterior of $T$ to be obtained subject to a normalizing constant: $p(T|X,Y) \propto p(Y|X,T)p(T)$.
The enormous number of trees that may be constructed from all but the smallest of data will render an exhaustive search of $p(T|X,Y)$ computationally intractable.
Subsequently, the normalization constant will not be obtainable nor will it be possible to identify which trees have with the highest posterior probability.
However, the posterior may still be explored using an MH algorithm to conduct a stochastic search.
Such a stochastic search will result in a Markov chain sequence of trees that converges towards trees with higher posterior probabilities and converges in distribution to the posterior $p(T | Y, X)$.
Chipman et al. (1998) note that their MH algorithm has a tendency to move rapidly towards a group of similar trees with high posterior probability proximate to the starting tree then remain in that vicinity exploring that group of trees with small local steps for many subsequent iterations of the algorithm.
Given a large enough number of iterations MH algorithms will move between posterior modes and explore the entirety of the trees possible but there is no guarantee about how many iterations a MH algorithm must be run for in order to achieve such a complete exploration.
In light of these considerations, Chipman et al. (1998) recommend comparing the results of multiple runs of their MH algorithm each originating from different starting values (origin trees).
Chipman et al. (1998) recommend both multiple restarts of their MH algorithm from the single root node tree, citing high initial variability in the direction in which their MH algorithm will proceed often leading such restarts to converge on quite different trees, and multiple restarts from start values selected from interesting intermediate trees from previous runs of their algorithm or trees identified by other methods (e.g. bootstrap bumping \citep{Tibshirani1999}).
Such multiple restarts of the MH algorithm for fitting Bayesian CARTS will result in a range of selected trees which can either be model averaged or chosen between depending on the goals of the particular analysis.
Aids for selecting the trees to include in the model averaging or for selecting a single tree could include the residual sum of squares of the resulting tree (either alone or constituent in a cost complexity metric) or plots of the observed likelihood of the trees $p(Y|X,T)$ against the number of terminal nodes of the same trees as a guide to the cost - complexity compromising being struck.

\subsubsection{Bagged Trees} 
Perhaps the simplest elaboration of CART comes from applying it within a bootstrapping procedure.
The term `bagging' was created by compressing the term `bootstrap aggregation' and refers to taking an average of the set of predictions obtained from applying the same model fitting procedure to a collection of bootstrap samples of some data.
Bagged Trees are reviewed in Hastie et al. (2009) who make the following observations regarding properties of Bagged Trees that would be pertinent to the application this technique to data in which linear and non-linear effects of covariates are expected to be important and collinearity is extant among covariates.
The average of many regression trees fitted to bootstrap resamples can better approximate linear and non-linear trends in the data than a single regression tree.
This stems from how the average of many such stepwise approximations to a linear (or non-linear) relationship, many of which will differ slightly having been fitted to different bootstrap resamples of the data, the will form a much better approximation to this relationship than any of the individual constituent stepwise approximations.
Collinearity among the covariates can lead to high variance among regression trees fitted to replicate data which bagging can smooth out in the hope of thereby obtaining a more generally applicable model.
However bagging regression trees will not improve the bias in estimation relative to that associated with a single regression tree fit.
This improvement in prediction due to reduction in variance from bagging comes at the cost of the interpretability of a single tree.
This is occurs since bagged regression trees are an average of the predictions of many such binary trees that have different geometries and as such the simple binary, branching nature of a single regression tree is sacrificed.

\subsubsection{Random Forests}
Like bagging, random forests \citep{Breiman2001} also involve averaging the predictions of a set of binary tree based models such as regression trees.
Furthermore, both bagging and random forests build this set of tree based models from a set of bootstrapped samples of the data.
Random forests elaborate on bagged trees by building `de-correlated' trees.
These de-correlated trees are propagated by choosing each binary division based on a randomly selected subset of the potential covariates.
Repeating this tree propagating process many times, each on a different bootstrap resample of the data results in a set of `de-correlated' trees that are then averaged to obtain a final prediction.
As such random forests share the advantage of bagged trees over a single regression tree in that they may better approximate linear and non-linear trends in the data while still being able to model complex interactions between covariates.
Random forests frequently perform synonymously to boosted trees and are easier to train and tune \citep{Hastie2009}.

\subsubsection{Boruta All Relevant Variable Selection } 
Random forests have also been taken as a starting point for further methodological elaborations and refinements.
One such technique is Boruta all relevant variable selection \citep{Kursa2010}, hereafter BARVS (our acronym).
The essence of the BARVS method is the recursive process of fitting a random forest then assessing which of the covariates utilised in this random forest made a sufficient contribution to the predictive performance of this random forest to warrant retention.
The random forest is then refitted, this time using only the covariates deemed worth retaining in the previous iteration. 
This process in then repeated until a random forest is fitted in which all covariates utilised are deemed to have made sufficient contribution to the predictive performance of the random forest to warrant retention.
Kursa et al. assess the contribution of a covariate to the predictive performance of the random forest via a $Z$ score (though note that $ Z \nsim N(0, 1) $ ).
The Boruta $Z$ score for a covariate in a particular random forest relates to the loss of accuracy of prediction resulting from the random permutation of the values of that covariate among observations.
The Boruta $Z$ score for a covariate in a random forest is calculated from these losses of predictive accuracy from all the constituent regression trees that utilized that covariate.
In particular, the Boruta $Z$ score is calculated by dividing the mean of these losses in predictive accuracy by the standard deviation of these losses (hence the choice of name).
The BARVS algorithm runs approximately as follows.
Firstly, permuted copies of all covariates currently under consideration for inclusion in the random forest are added to this set of considered covariates and a random forest is fitted to these composite data.
The $Z$ scores are then calculated for all of covariates, including the permuted copies of the actual covariates.
The maximum $Z$ score among the permuted covariates is then identified and two sided $t$ tests are performed to test the equality of the $Z$ score for each actual covariate against this maximum $Z$ score among the permuted covariates.
Next, all actual covariates that have $Z$ scores significantly less than this maximum $Z$ score among the permuted covariates are discarded from the set of considered covariates and the current set of permuted covariates is also discarded. 
This procedure is then repeated until all remaining actual covariates have $Z$ scores significantly greater than the maximum $Z$ score among the permuted covariates created at that iteration.
The random forest constructed from these exclusively relevant covariates is then retained as the final solution.

\subsubsection{Boosted Trees}
Boosting is a well regarded technique that may be applied to a variety of models including binary tree based models \citep{Hastie2009}.
Boosted regression trees have been introduced in a manner accessible to quantitative scientists by Elith et al. (2008).
Akin to bagged trees and random forests, boosted trees incorporate a sequence of binary trees (such as regression trees) fitted to sequentially modified versions of the original data, the final prediction being drawn from a combination of the predictions of this sequence of trees.
In the case of boosted trees, the sequential modifications to the original data take the form of weightings calculated from the results of the tree fitted in the previous step.
Boosted regression trees implement a type of functional gradient descent designed to minimise a loss function that quantifies the loss in predictive accuracy resulting from an imperfect model fit \citep{Elith2008}.
The first tree is fitted to the original data by maximising the reduction in the value of the loss function relative to that from a single node tree.
The second tree is fitted to the residuals from the first tree but the predictions from this second step in the fitting process are the result of combining the predictions first and the second tree.
The third tree is then fitted to the residuals from the combined predictions of the first and second trees and so on.
In this manner boosting sequentially focuses the model fitting on the observations that are difficult to explain.
Boosting is typically conducted for as many iterations as is computationally feasible and the final prediction from the ensemble of boosted regression trees is a type of weighted average of the predictions from all the constituent trees.
Predictions from boosted regression trees have increased stability and accuracy compared to those from a single regression tree model.
Furthermore, the introduction of some stochasticity into the boosting algorithm via the inclusion of a bagging step can further improve the accuracy of the predictions and mitigate the effects overfitting \citep{Friedman2002}.
Boosted trees like bagged trees and random forests involve the sacrifice of the interpretability of a single regression tree for better predictive performance.
However, a relatively straight forward metric of covariate importance exists \citep{Friedman2001, Friedman2003} which scores covariates based on the frequency with which each defined a binary division and weights these scores proportionally to the improvement in the model fit that resulted from the inclusion of the associated division.
These scores are averaged across all the trees in the boosting sequence and the resulting scores are scaled to sum to 100 for ease of interpretation.
Unlike a single regression tree, boosted regression trees can easily model linear relationships, non-linear relationships and relationships that include step-like discontinuities \citep{Elith2008}.
Unlike bagged trees and random forests, boosted regression trees reduce both the variance and bias associated with predictions \citep{Elith2008}.

\subsubsection{Cubist}
Cubist (\url{https://www.rulequest.com/cubist-info.html}) fits predictive models developed from Quinlan’s M5 model tree method \citep{Quinlan1992}.
Quinlan's M5 method functions by creating a binary tree structure which is then pruned to reduce tree complexity without greatly reducing the overall fit of the tree to the data.
Where this pruning converts former interior nodes of the tree into terminal nodes by collapsing the tree structure below them, MLR models are fitted to each of the subsets of the data thus defined.
Each MLR model selects covariates from the set of covariates that were previously used to define the tree structure that was below (child nodes of) this current node.
The Cubist method extends the M5 method by incorporating a boosting step.

\subsubsection{Bayesian Treed Regression}
The Bayesian implementation of CART \citep{Chipman1998} has been extended to fit MLR models (rather than simple intercept only models) to the subsets of the data defined by the terminal nodes of the associated binary tree in a framework the creators dubbed Bayesian treed regression \citep{Chipman2002}.
One motivation for such a model formulation being the scenario whereby different covariates are most useful for predicting the response in different subsets of the data.
The end result is somewhat akin to C5 model trees \citep{Quinlan1992} but the method of attaining such a fit is very different, the model being formulated under the Bayesian paradigm.
Much like Bayesian CART, Bayesian treed regression utilises an implicitly specified tree prior and a stochastic search.
The Bayesian formulation and stochastic search enable different tree geometries and MLR models in the terminal nodes of these trees to be explored in addition to different error variances to be modelled for each terminal node.
The particulars of the model formulation are as follows.
For each terminal node of the tree $T$, the response observations, $Y$, that are members of this node are assigned a parametric model.
In this manner there is a separate parametric model for the response observations contained in each of the unique terminal nodes of the tree.
In particular, the distribution of the elements of the response vector $\mathbf{Y}$ that are members of the $i^{\text{th}}$ terminal node of the tree $T$ are modeled conditional upon the associated covariate values by the parametric model $Y|x \sim f(y|x,\theta_i)$ where $\Theta = (\theta_1, ... , \theta_b)$.  
This contrasts with a Bayesian CART where the distributions of the response observations in each terminal node are not modeled as conditional upon the associated observations of the covariates $x$ there.
Where CART and Bayesian CART models utilise a collection of stepwise functions to approximate correlations between the response and the covariates via the binary tree structure Treed regressions may be thought of as a collection of piecewise MLR models.
As such Bayesian treed regression models would model linear or non-linear relationships between covariates and the response much more parsimoniously than Bayesian CART models as Bayesian treed models have the facility to invoke MLR in the terminal nodes.
This facilitates the transfer of complexity in Bayesian treed models from the tree geometry to the terminal node MLR models.
As such, one may reasonably expect shallower more readily intelligible trees to be coupled with the terminal node MLR models in Bayesian treed regression as compared to the more deeper, less readily intelligible trees that are coupled with terminal node single parameter models in Bayesian CART.
This may be thought of as the binary tree component of the Bayesian treed models capturing the major subsets of the data and the associated MLR models describing the nuances thereof.
\newline
\newline
Each fit of a Bayesian treed model is uniquely defined by $(\Theta, T)$ and Chipman et al. (2002) outline how the posterior distribution of these $(\Theta, T)$ may be explored via stochastic search with the aid of a Metropolis Hastings algorithm.
As with their Bayesian CART, Chipman et al. (2002) recommend comparing the fits to which multiple restarts of the stochastic search converge so as to better explore the posterior rather than running a single stochastic search for a long time.
They make this recommendation in light of the noted tendency of their MH explorations to converge on local maxima in the posterior then explore the vicinity of that maxima for many iterations.
Chipman et al. (2002) note that their approach is amendable to both Bayesian model selection, should a single model be desired, and Bayesian model averaging (e.g. by posterior or likelihood weighting of the iterations of the stochastic search) should better predictive performance be desired.
When model averaging is elected, Chipman et al. suggest model averaging only the better fitting models.

\subsection{An Introduction to the Multiple Linear Regression Based Models Considered this Work}
\label{SS:MLR}
Multiple Linear Regression (MLR) predicts observations of the response $y_i$, $i = 1, ..., n$, from a linear combination of products of observations of the covariates $x_{i,j}$, $j = 1, ..., p$, and the associated coefficient estimates $\hat \beta_j$ plus an intercept term $\hat \beta_0$ as per Equation \ref{Eq:MLR}:
\begin{equation}
\label{Eq:MLR}
\hat y_{i} = \hat \beta_0 + \sum_{j=1}^p \hat \beta_j x_{i,j}
\end{equation}
In MLR, non-linear effects of covariates upon the response may be modelled via inclusion of additional covariate terms formed as single term polynomials constructed from the original covariates.
Interactions between covariates in their effects upon the response may be modelled via including in the regression additional covariate terms constructed by taking products of the covariates constituent to the interaction in question.

\subsubsection{MLR Maximum Likelihood}
When a MLR model is fitted by ordinary least squares (OLS) likelihood maximisation the vector of coefficient estimates $\mathbf{\hat \beta}$ is obtained from the design matrix $X$ as per Equation \ref{Freq:Beta.Hat}.
\begin{equation}
\label{Freq:Beta.Hat}
\mathbf{\hat \beta} = \bigl(  X^T X  \bigr) ^{-1} X^T \mathbf{y}
\end{equation}
As such ordinary least squares cannot estimate MLR fits for scenarios where the number of covariates, $p$, exceeds the number of observations, $n$, (the $p>n$ or ultrahigh dimensional scenario).
Even in situations where the number of covariates is less than the number of observations, consideration of non-linear terms for each covariate and the $\binom{p}{k}$ possible order $k$ interaction terms can lead to the number of considered covariate terms grossly exceeding the number of observations and OLS fitting no longer being possible.
Furthermore, fitting MLR by OLS is ill advised when collinearity exists among the covariates \citep{Belsley1980}.

\subsubsection{MLR with Information Criterion Based Variable Selection}
When the number of covariate terms desired to be considered for use in predicting the response exceeds the number of observations, OLS is often incorporated into a model comparison framework that fits and compares numerous models that utilise subsets of the available covariates such that $p < n$.
This comparison is often effected via an information criterion such as the Akaike Information Criterion (AIC) \citep{Akaike1974,Venables2002,Konishi2008} which assigns models a score that rewards goodness of fit to the data while penalizing model complexity.
Where it is computationally feasible to do so all possible models that may be constructed from a particular set of covariates with a particular number of observations may be fitted via an exhaustive search procedure and compared, for instance in terms of information criterion values.
Where the computational burden of an exhaustive search is deemed too great, a popular choice is to adopt some greedy algorithm based approach to search for optima in the information criterion values that accompany the set of possible models without having to fit all these models.
Stepwise variable selection methods are examples of such techniques.
It should be noted that in their base forms both stepwise variable selection and exhaustive searches still utilise OLS fitting procedures and as such are both ill advised in the presence of collinearity among the covariates.

\subsubsection{MLR Bayesian}
The Bayesian approach may also be used to fit MLR models (see for example Gelman et al. 2004).
Furthermore, the Bayesian framework may be used to accomplish variable selection under the ultrahigh dimensional scenario by creating prior distributions that give each regression coefficient a high probability of being zero \citep{Gelman2004} as performed in the Bayesian variable selection method Spike and Slab priors \citep{Mitchell1988, Geweke1996, George1997}.
A simple Bayesian formulation of an MLR with independent and non-informative priors on all coefficients may be adversely affected by collinearity among covariates.
Collinearity among covariates will lead to high posterior variance of the associated coefficient estimates \citep{Gelman2004} and subsequently slow Markov Chain Monte Carlo convergence.
Bayesian analogues to shrinkage techniques for mitigating the undesirable effects of modeling with data that includes collinearity among the covariates are outlined along side their penalized likelihood based counterparts in the following section.

\subsubsection{MLR Penalization / Shrinkage}
One of the dangers when conducting MLR based modelling with covariates among which collinearity exists is that highly correlated pairs of covariates can be assigned arbitrarily large magnitude positive and negative coefficients that effectively negate each other.
One action that may be taken to mitigate or at least control this effect is to impose a penalty upon the combined magnitude of the coefficients as part of the fitting process \citep{Hastie2009}.
A simple choice for such penalty functions is to take the $L_\gamma$ norm of the regression coefficient vector $\mathbf{ \beta}$, for some value of $\gamma$, and search for the $\mathbf{ \hat \beta}$ that minimizes the sum of the residual sum of squares and this norm of the regression coefficients as per Equation \ref{eq:L.gamma.pen.ls}.
\begin{equation} \label{eq:L.gamma.pen.ls}
\mathbf{\hat \beta}^{L_\gamma}  = \argmin\limits_{\mathbf{\beta}} \{ \sum\limits_{i = 1}^n (y_{i} - \beta_0 - \sum\limits_{j = 1}^px_{ij} \beta_j)^2 + \lambda \sum\limits_{j = 1}^p | \beta_j |^\gamma  \},\ \ \gamma > 0
\end{equation}
Solving Equation \ref{eq:L.gamma.pen.ls} with $\gamma = 2$ yields a Ridge Regression estimate equivalent to that obtained by solving Equation \ref{Eq:Ridge}.
\begin{equation}
\label{Eq:Ridge}
\mathbf{ \hat  \beta}^{\text{ridge}} = \argmin\limits_{\mathbf{\beta}} \{ \sum\limits_{i = 1}^n (y_{i} - \beta_0 - \sum\limits_{j = 1}^px_{ij} \beta_j)^2 + \lambda \sum\limits_{j = 1}^p \beta_j^2 \}
\end{equation}
Ridge regression shrinks all coefficients towards zero and thus does not perform variable selection and is not useful for regression in the ultrahigh dimensional situation.
Interestingly, ridge regression is equivalent to Bayesian MLR with an exchangeable normal prior distribution on the coefficients \citep{Gelman2004}.
Using $\gamma = 1$ in Equation \ref{eq:L.gamma.pen.ls} yields an $L_1$ penalized least squares estimate also known as the Least Absolute Shrinkage and Selection Operator (LASSO) \citep{Tibshirani1996}.
More complex choices for the penalty function than the $L_\gamma$ norm in Equation \ref{eq:L.gamma.pen.ls} are used in penalized least squares techniques such as adaptive LASSO \citep{Zou2006}, Smoothly Clipped Absolute Deviation (SCAD) \citep{Fan2001} and Minimax Concave Penalty (MCP) \citep{Zhang2010}.
Of these techniques $L_\gamma$ penalization is perhaps the most easily applied.
Solving Equation \ref{eq:L.gamma.pen.ls} for any $\gamma < 2$, will shrink the coefficient estimates for some covariates to zero exactly (how many depends on value of tuning parameter $\lambda$) thereby performing variable selection in addition to penalized estimation \citep{Ahmed2014}.
As such $L_\gamma$ penalized estimation with $\gamma < 2$ is applicable in scenarios where where the number of potential covariates exceeds the number of observations ($p > n$) and collinearity exists among the covariates.
The absolute value in an $L_1$ penalized estimate requires a computational solution initially provided by quadratic programming \citep{Tibshirani1996} and more recently by the more computationally efficient Least Angle Regression (LAR) algorithm \citep{Efron2004}.
An estimate of the LASSO solution may also be obtained from the posterior mode estimates of a Bayesian MLR with independent and identical, Laplace (double exponential) priors on the regression coefficients \citep{Park2008}.

\subsection{Comparing Multiple Linear Regression Based Techniques and Binary Tree Based Technique} 
\label{SS:MLRcfBT}
The comparison of the properties of Multiple Linear Regression (MLR) and Classification And Regression Tree (CART) has most relevance to this work as a foundation to inform the comparison of the various modifications of each technique that would have better suited the coupling of the particular characteristics of the data from our case study and our objective for the analysis of these data.
We consider the defining characteristics of our case study data to be: (1) more potential covariate terms than observations (the $p > n$ or ultrahigh dimensional situation for variable selection)  (2) a high degree of collinearity among the potential covariate terms and (3) suspected importance of non-linear effects of covariates and interactions of covariate effects.
The primary objective with our case study analysis was the construction of a model for covariate assisted interpolation of the response.
Our case study also had the additional context of the modest computational resources provided by one mid-range laptop and our desire for an easily interpretable predictive mechanism.
\newline
\newline
A plethora of BT based approaches exist today that are variously modifications of and elaborations upon the Classification And Regression Trees \citep{Breiman1984} (hereafter CART) framework.
We confine ourselves here to considering a subset of these that appeared appropriate for the case study data and objectives namely: Bayesian CART \citep{Chipman1998}, bagged regression trees \citep{Breiman1996}, random forests \citep{Breiman2001}, boruta all relevant variable selection \citep{Kursa2010} (an elaboration upon random forests), boosted regression trees \citep{Friedman2002}, cubist \citep{Quinlan1992} and Bayesian treed regression \citep{Chipman2002} (an elaboration upon Bayesian CART).
Readers unfamiliar with these BT based techniques are referred to Section \ref{SS:BT}.
A similar diversity of modifications to and elaborations upon MLR exist though we confine ourselves here to considering the base form and the modification thereof that seem appropriate to the defining features of our case study data.
Namely, we consider shrinkage modified MLR for its relevance to the situation where collinearity exists among the covariates with particular attention to LASSO style shrinkage for its additional relevance to the situation where the number of covariates exceeds the number of observations of the response (see Section \ref{SS:MLR} for an introduction to these techniques).
We consider both LASSO for MLR fitted under the Bayesian paradigm \citep{Park2008} and LASSO implemented through likelihood penalization as fitted by the LAR algorithm \citep{Efron2004}.
We consider each of these techniques in light of the key characteristics of our case study data and our objective of building a model to interpolate the response.
This allows us to narrow down the choice of methods further and make a final decision.
This comparison is also summarised in Table \ref{tab:MLR_cf_BT_Long_Tab}.
\newline
\newline
We commence this consideration with perhaps the most widely know technique introduced above, MLR.
A MLR model cannot be fitted by Ordinary Least Squares (OLS) based likelihood maximisation when the number of covariates desired to be included in the model exceeds the number of observations (see Section \ref{SS:MLR}). 
The large number of potential covariate terms and suspected importance of non-linear effects and interactions of covariate effects meant that conducting exhaustive search variable selection on these data was not computationally feasible.
Under such scenarios, a common option has been to apply some deterministic variable selection procedure that optimises an information criterion.
Stepwise variable selection with the Akaike Information Criterion (AIC) \citep{Akaike1974,Venables2002, Konishi2008} has been a popular choice in this regard having been available in R via the step function \citep{R2015} through numerous release cycles.
However, with linear regression, correlations between potential covariates can be a cause for concern with such techniques that rely on ordinary least squares model fitting (which in the case of linear regression is also maximum likelihood model fitting). 
The undesirable effects of correlations among covariates in a MLR model have been explained in the comprehensive book \citep{Belsley1980}.
While Bayesian methods would allow a MLR model to be fitted with more covariates than observations the high degree of collinearity among some pairs of covariates would render the associated pairs of coefficients poorly defined which in turn could complicate convergence of the MCMC iterations upon a fit to these data.
In the presence of collinearity among covariates some form of shrinkage is advisable for preventing highly correlated covariates being assigned arbitrarily large magnitude but opposite signed coefficients that all but cancel each other out due to the high correlation of the associated covariates.
Coefficient shrinkage is performed by the popular Penalized Least Squares (PLS) family of techniques \citep{Ahmed2014} which forms a subset of the larger family of coefficient shrinkage themed modifications of MLR.
We considered shrinkage techniques including Ridge regression \citep{Hoerl1970}, LASSO \citep{Tibshirani1996}, LASSO fitted by LAR \citep{Efron2004} and the Bayesian LASSO \citep{Park2008}.
Of particular interest were both LASSO for MLR fitted under the Bayesian paradigm \citep{Park2008} and LASSO implemented through likelihood penalization as fitted by the LAR algorithm \citep{Efron2004} since LASSO is appropriate to both the situation where the number of covariates exceeds the number of observations and the situation where substantial collinearity exists among the covariates.
LASSO conducts shrinkage in a manner that shrinks some coefficients to zero exactly which in effect excludes them from the model and as such is useful for conducting shrinkage and variable selection simultaneously.
A LASSO modified MLR fit may be obtained via the deterministic and computationally efficient LAR algorithm \citep{Efron2004} which adds or removes the covariate that best optimises its criteria at each iteration.
Alternatively, a LASSO modified MLR estimate may be obtained by fitting a Bayesian MLR with independent and identical Laplace priors on all the coefficients \citep{Park2008}.  
Such an exploration, utilising multiple MCMC chains with a variety of starting values, could well conduct a more thorough exploration of the possible LASSO fits of MLR models that could be constructed from the potential covariates than would be conducted by the deterministic LAR algorithm.
Though we note that were non-linear effects and all possible pairwise interactions considered as potential covariate terms, the computational burden involved in obtaining convergence of all chains could be considerable simply due to the number of coefficients to be estimated.
\newline
\newline
In contrast to MLR, CART may be run in its base form on ultrahigh dimensional data given sufficient computational resources.
With more covariates than observations it would be vital to enforce a minimum number of observations per terminal node at which to commence pruning but if this were done there would be no a priori barrier to the implementation of this technique on such data.
Were CART models being fitted within the Bayesian paradigm the tree prior would need to be parameterised to perform a similar function guiding the posterior away from deep trees with low numbers of observations per terminal node.
Furthermore, the collinearity present among covariates would not prove a barrier to the implementation of CART (here regression trees as we have a continuous response) via the standard algorithms as even small differences among covariates would be sufficient for one covariate to define a binary partition of the response that reduces some loss metric more than the other covariate.
The 63 covariates of our case study analysis, while not prohibitively excessive, would likely lead to some computational burden when fitting Bayesian CART models to these data as the stochastic searches necessary may take many iterations to converge.
We postulate this slow convergence due to the collinearity among some covariates and also due to the number of covariates considered.
Given the tendency of Bayesian CART to converge upon a local maxima and remain there moving locally for many MCMC iterations thereafter \citep{Chipman1998} the greater the number of chains with a diversity of starting trees that could be run the greater the conviction one could have that a good enough exploration of the posterior had been conducted to have at least identified some trees that were more than just local maxima of extremely small regions of the posterior.
As such a good case exists for devoting significant computational resources to fitting Bayesian CART.
Suspected interactions of covariates and non-linear effects of covariates upon the response are a more interesting issue in the context of regression trees.
Regression trees by their nature can model complex interactions between covariates and this is a strength they possess relative to MLR.
However, as a single regression tree is essentially a collection of stepwise predictions for the response across mutually exclusive partitions of covariate space, deep trees will be necessary to approximate even a simple linear relationship between a covariates and the response well.
The limited number of response observations available in our case study and large number of covariates of which several or more may well have non-linear relationships with the response means that trees of sufficient depth to describe multiple non-linear relationships would not be possible with sufficient observations in each terminal node to formulate reliable estimates of the response means there.
Subsequently, some form of averaging of multiple trees would be advisable as the average of many slightly different stepwise functions can well approximate a linear or non-linear relationship even when the constituent binary trees are quite shallow. 
The trees thus averaged could be the trees that scored above some threshold posterior probability or trees fitted to multiple re-samples or re-weightings of the full set of observations.
\newline
\newline
Fitting regression trees to numerous bootstrap re-samples of the data would be the simplest option for generating multiple regression trees to model average in order construct better approximations to linear and non-linear relationships between covariates and the response than would be possible with a single binary tree.
This technique is referred to as bootstrap aggregated trees or bagged trees for short \citep{Breiman1996}.
Furthermore, averaging the predictions from numerous shallow trees each fitted to a different bootstrap re-sample of the data could also allow the various linear and non-linear relationships between different covariates and the response to be incorporated into the model despite the small sample size.
If different covariates were most important to different subsets of the response, different trees could result from fitting a regression tree to each bootstrap re-sample and thus this diversity of relationships could be incorporated into the bootstrap aggregation.
In this manner, even though the response sample size limits the depths of trees possible, numerous linear and non-linear relationships between covariates and the response could be incorporated into the predictions through the different shallow trees constituent to the model averaging.
As such, bagged trees could incorporate approximations of far more linear and non-linear relationships into predictions of the response than it would be possible to utilise via a single binary tree fitted to the same data.
As predictions from bagged trees are constructed from the predictions of numerous individual regression trees the same ability to operate on data with high collinearity among the covariates exists for bagged regression trees as does for single regression trees.
Similarly, bagged regression trees retain and even enhance the strength of single regression trees at modelling complex interactions between covariates upon their effect on the response.
However, these advantages come at the cost of the easily interpretable structure of a single regression tree.
Much the same assertions may be made for random forests.
\newline 
\newline
Random forests \citep{Breiman2001}, like bagged trees, model average predictions from a set of regression trees derived from bootstrap re-samples of the data and thus have similar advantages to bagged trees over single regression trees at approximating linear and non-linear relationships between covariates and the response.
Random forests differ from bagged trees in the manner in which regression trees are fitted to the bootstrap re-sample of the data.
At each binary partition in such a tree in a random forest, only a randomly selected subset of the potential covariates are made available to the algorithm to define the binary partition.
Random forests may be better than bagged trees at approximating different linear and non-linear relationships between the covariates and the response with a limited number of observations due to the combination of bootstrapping and selecting from random subsets of covariates at each partition resulting in an broader range of shallow trees to aggregate than bootstrapping alone.
Random forests, being composed of many individual regression trees, should not be adversely affected by collinearity among covariates and should still be able to model complex interactions.
\newline
\newline
Boruta all relevant variable selection \citep{Kursa2010} is an elaboration of the random forest method.
As such, it shares the advantages of bagged trees and random forests over single regression trees for approximating linear and non-linear relationships between covariates and the response and, like all binary tree based methods, is able to model complex interactions among covariates in their effects upon the response.
However, collinearity among the covariates may pose a problem for the application of boruta all relevant variable selection.
The problem stems from the boruta algorithm being vulnerable to discarding useful covariates when these covariates are highly correlated with one or more other covariates.
If highly correlated covariates were supplied to a random forest algorithm and these covariates contained useful information for predicting the response these covariates could well be used interchangeably throughout particular regression trees for defining binary partitions of the response.
Thus the loss in predictive accuracy that would result from permuting the values of just one of these correlated covariates and using it in place of the actual covariate would likely be less than would result if this particular covariate had been used in all binary partitions of the tree where the partition was defined based on one of the other covariates with which this covariate was highly correlated.
Subsequently, the importance of this particular covariate to the predictive accuracy of the regression tree could be underestimated to the extent that the Boruta all relevant variable selection algorithm would discard this covariate from the set of covariates worth retaining.
Furthermore, if the correlated covariates in question were sufficiently correlated that they were being used interchangeably throughout the regression tree then they could all dilute the estimated importance of the others in this manner such that none of these correlated covariates were selected for retention by the boruta all relevant variable selection algorithm despite all containing much the same important information for predicting the response and thus it being well worth retaining one of them.
\newline
\newline
Boosting iteratively refits a model to data that is re-weighted at each iteration to emphasize the observations that were poorly predicted by the model fitted in the previous iteration.
The predictions from this sequence of models are then combined to produce the final prediction.
Where these models are regression trees the procedure is referred to as boosted regression trees \citep{Friedman2002}.
Thus the predictions constructed from a sequence of boosted regression trees will have similar advantages to those from bagged trees and random forests over a those from a single regression tree model.
Such aggregated predictions will better approximate linear and non-linear relationships than the predictions from a single shallow regression tree given the small number of response observations and numerous potential non-linear and interacting effects of covariates expected to exist in our data.
Furthermore, by its very nature boosting will produce a variety of trees each fitted to re-weightings of the data which emphasize a different subset of the response observations.
Thus, where the small sample size would hinder a single regression tree capturing the suspected importance of multiple non-linear relationships between covariates and the response, different trees in the sequence of boosted trees will describe different relationships between covariates and the response that are found to be important to observations up weighted at that iteration.
Since the ensuing predictions from all such trees will then be aggregated to produce the final predictions all these different identified relationships will be combined into the predictions of the response.
Again, similar to the above approaches that aggregate sequences of regression trees to produce predictions, collinearity among the covariates should not greatly hinder the regression tree fitting process constituent to fitting boosted regression trees as differences will exist among even quite collinear covariates sufficient to choose between them when they are useful for defining a binary partition of the response.
Furthermore, still being based on binary trees, boosted regression trees will be able to model complex interactions between covariates in their relationships with the response.
\newline
\newline
The cubist method \url{https://www.rulequest.com/cubist-info.html} is in essence an extension of the M5 or model tree approach \citep{Quinlan1992} that incorporates a boosting step.
As such the model that is fitted to the iteratively re-weighted data at each iteration of the boosting algorithm starts as a regression tree which is then iteratively pruned back by collapsing parent nodes to the current terminal nodes and fitting MLR models to the subsets of the data defined by these newly created terminal nodes.
Each MLR model fitted to the data contained in a newly created terminal node uses as potential covariates all the covariates that defined the binary partitions that were collapsed to create this new terminal node.
As such, collinearity among the covariates could lead to poorly defined coefficients in these terminal node MLRs as discussed above (assuming these covariates were also collinear in the subset of observations contained the terminal node in question) unless some shrinkage based fitting was conducted there.
This model structure would allow for very flexible description of linear and non-linear relationships between covariates and the response as not only are different stepwise predictions being averaged but constituent in this averaging are also the predictions for various MLR fits.
Similarly, being based upon a binary tree structure complex interactions may be modeled implicitly by this method.
\newline
\newline
Bayesian treed regression \citep{Chipman2002} is superficially similar to Quinlan's M5 in that it fits MLR models to subsets of the data contained in the terminal nodes of a binary tree.
Bayesian Treed Regression, however, is fitted under the Bayesian paradigm via stochastic search as an elaboration of a Bayesian CART model \citep{Chipman1998}.
The intricate model structures permitted by fitting MLR models in the terminal nodes of binary trees is attractive for data in which complex combinations of linear, non-linear and interaction effects of covariates upon the response are suspected to be important.
Some form of shrinkage would be advisable in the terminal nodes to mitigate the concerns of conducting MLR with collinear covariates.
As has been discussed above LASSO may be implemented within the Bayesian framework by placing Laplace priors on the regression coefficients \citep{Park2008}.
Our greatest concern associated with this method would be the computational burden inherent in conducting a good exploration of the posterior.
Collinearity among covariates would slow the convergence of stochastic searches and the shear breadth of possible models would require numerous chains to be run with a great variety of starting values in order to have any confidence whatsoever that a good exploration of the posterior had been conducted given the noted tendency of this algorithm to rapidly converge on local maxima in the posterior then remain in the neighbourhood of this maxima for many subsequent MCMC iterations \citep{Chipman2002, Chipman1998}.
The number of parameters in the stochastic search could be reduced if one were willing to let the binary tree portion of the model be the only form of allowing for potential interactions of covariates (i.e. consideration of the $\binom{63}{2}$ pairwise interaction terms in the MLR models could then be avoided).
Further reduction in the number of parameters in the stochastic search could be achieved by only allowing linear terms for each covariates in the terminal node MLRs and relying on model averaging of high posterior probability fits to account for non-linear effects through the averaging of the predictions from multiple combinations of stepwise functions with MLR fits.
Such an approach coupled with a tree prior parameterised to avoid deep binary trees would allow for a rich variety of linear, non-linear and interactions effects to be incorporated into the aggregated predictions.
\newline
\newline
Of the techniques considered, LASSO penalized MLR fitted via the LAR algorithm, random forests and boosted trees are here proposed as the techniques that were most suitable for our case study analysis (see Table \ref{tab:MLR_cf_BT_Long_Tab}).
These three techniques: were appropriate for application to data with characteristics like those of our case study, suited our objective of building models to interpolate the response, were straight forward to implement with existing software and seemed unlikely to be adversely computationally burdensome with the computational resources available.
Model-averaging of high posterior probability regression trees identified via the Bayesian CART method also appeared quite well suited to data and objectives like ours but would have required more effort to implement (tree priors require some work to parameterise so as to emphasize tree structures that maintain a minimum number of observations in all terminal nodes) and could well have proved quite computationally expensive with the computational resources available.
Model-averaging of high posterior probability fits from Bayesian treed regression also appeared very promising for data and objectives like ours provided some form of shrinkage could be implemented in the terminal node MLRs and the 63 covariates were considered only as linear main effects.  
However, we note that the implementation of shrinkage in the terminal nodes would not be a trivial task and that both these techniques seemed likely to be quite computationally intensive with data like ours.
The choice of an easily implemented technique that was appropriate to our data and computationally efficient was thus reduced to a choice between LASSO penalized MLR fitted via the LAR algorithm, random forests and boosted trees.
\newline
\newline
We have elected to use LASSO modified MLR fitted via the LAR algorithm in our case study analysis. 
Model-averaging the predictions from the LASSO solutions obtained from LAR executions within a cross validations scheme yielded an aggregate estimate in a manner conceptually similar to the manners in which random forests, bagged trees or boosted trees aggregate predictions from the same model structure fitted to variations on the same data set.
Another useful consequence of using a cross validation based approach was that we were able to estimate the shrinkage parameter for the LASSO fits ($\lambda$ in Equation \ref{eq:L.gamma.pen.ls}) via cross validation.
The motivation for this approach was an attempt to fit models that would perform well at interpolation.
For the purposes of building models for interpolation, LASSO modified MLR fitted via the LAR algorithm was a defensible choice of method from a set of good options.
Our preference for LASSO modified MLR fitted via LAR was the result of a secondary interest in which non-linear terms for covariates and which particular pairwise interactions were most useful for predicting the response.
This was more readily apparent from LASSO modified MLR fits where each covariate term has a coefficient that has been either shrunk to zero, effectively excluding the covariate term from the model, or assigned a value.
When employed within a cross validation scheme this translated into frequencies of selection of these specific covariate terms which were still easy to interpret.
In contrast, whether the overall role of a covariate within the aggregated estimate from random forests, bagged or boosted trees was closer to linear or non-linear (and if non-linear what manner of non-linear) would have been harder to judge from the results of such a fit.
This ease of interpretability of the LASSO modified MLR came at a cost of having to recenter and rescale (to mean zero and magnitude one) all covariates in each training each set (a requirement of the LAR algorithm \citep{Efron2004}) and mirror those transformations on each associated validation set  whereas this would not have been necessary for binary tree based techniques.
\clearpage
\begin{sidewaystable}
\caption{A summary of the appropriateness of various MLR and CART based methods for modelling data of particular characteristics.}
\label{tab:MLR_cf_BT_Long_Tab}
\begin{tabular}{llll}
Method                                 & Data Characteristic & Method Pro & Method Con \\
\hline
\hline
MLR (Maximum                           & ultrahigh      & & insufficient degrees of freedom \\
Likelihood)                            & dimensionality & &  \\
 & & & \\
                                       & collinearity   & & coefficients poorly defined \\
 & & & \\
                                       & non-linearity  & models directly with polynomial terms & consumes degrees of freedom \\
 & & & \\
                                       & interactions   & models directly with product terms & consumes degrees of freedom \\
 & & & \\
MLR                                    & ultrahigh      &  & computationally expensive \\
(Bayesian)                             & dimensionality &  & \\
 & & & \\
                                       & collinearity   &  & coefficients poorly defined,   \\
 & & & computationally expensive\\
 & & & \\
                                       & non-linearity  & models directly with polynomial terms & computationally expensive \\
 & & & \\
                                       & interactions   & models directly with product terms & computationally expensive \\
 & & & \\
\end{tabular}
\end{sidewaystable}

\clearpage

\begin{sidewaystable}
\begin{tabular}{llll}
Method                                 & Data Characteristic & Method Pro & Method Con \\
\hline
\hline
MLR + Ridge                            & ultrahigh      &  & shrinks all coefficients,  \\
Regression         & dimensionality &  & does not perform variable selection\\
 \citep{Hoerl1970} & & & \\
                                       & collinearity   & constrains coefficients  & \\
 & & & \\
                                       & non-linearity  & models directly with polynomial terms & computationally expensive \\
 & & & \\
                                       & interactions   & models directly with product terms & computationally expensive \\
 & & & \\
MLR + LASSO via                        & ultrahigh      & performs variable selection & computationally inefficient \\
quadratic programming  & dimensionality & &  \\
\citep{Tibshirani1996} & & & \\
                                       & collinearity   & constrains coefficients &  \\
 & & & \\
                                       & non-linearity  & models directly with polynomial terms & computationally expensive \\
 & & & \\
                                       & interactions   & models directly with product terms & computationally expensive \\
 & & & \\
\end{tabular}
\end{sidewaystable}

\clearpage

\begin{sidewaystable}
\begin{tabular}{llll}
Method                                 & Data Characteristic & Method Pro & Method Con \\
\hline
\hline
MLR + LASSO                            & ultrahigh      & computationally efficient,  & \\
(LAR)                  & dimensionality & performs variable selection  & \\
\citep{Efron2004} & & & \\
                                       & collinearity   & constrains coefficients  & \\
 & & & \\
                                       & non-linearity  & models directly with polynomial terms & computationally expensive \\
 & & & \\
                                       & interactions   & models directly with product terms & computationally expensive \\
 & & & \\
MLR + LASSO                            & ultrahigh      & performs variable selection  & computationally expensive \\
(Bayesian)              & dimensionality &  & \\
\citep{Park2008}  & & & \\
                                       & collinearity   & constrains coefficients  & \\

 & & & \\
                                       & non-linearity  & models directly with polynomial terms & computationally expensive \\
 & & & \\
                                       & interactions   & models directly with product terms & computationally expensive \\
 & & & \\
\end{tabular}
\end{sidewaystable}

\clearpage

\begin{sidewaystable} 
\begin{tabular}{llll}
Method                                 & Data Characteristic & Method Pro & Method Con \\
\hline
\hline

CART                                   & ultrahigh      & enforces minimum $n$ per terminal node & \\
(Frequentist)       & dimensionality &   with stopping rule \& \/ pruning & \\
\citep{Breiman1984} & & & \\
                                       & collinearity   & not adversely affected  & \\
 & & & \\
                                       & non-linearity  & & requires deep tree \& large $n$ \\
 & & & \\
                                       & interactions   & may describe complex interactions & \\
 & & & \\

CART                                   & ultrahigh      & tree prior may be tuned to pull posterior & computationally expensive\\
(Bayesian)           & dimensionality &  towards trees with desired $n$ per terminal node & \\
\citep{Chipman1998} & & & \\
                                       & collinearity   &  & extends stochastic search convergence time \\
 & & & \\
                                       & non-linearity  & Bayesian model-averaging could & large $n$ for deep tree required \\
 & &  approximate well &   for single tree to approximate well \\
 & & & \\
                                       & interactions   & may describe complex interactions &  \\
 & & & \\
\end{tabular}
\end{sidewaystable}

\clearpage

\begin{sidewaystable}
\begin{tabular}{llll}
Method                                 & Data Characteristic & Method Pro & Method Con \\
\hline
\hline
Bagged Trees        & ultrahigh      & stopping rule \&$/$ pruning enforces minimum    & \\
 \citep{Breiman1996}                & dimensionality & $n$ per terminal node & \\

 & & & \\
                                       & collinearity   & not adversely affected & \\
 & & & \\
                                       & non-linearity  & approximates well by averaging many shallow trees &  \\
 & & & \\
                                       & interactions   &  & may describe complex interactions \\
 & & & \\
Random Forests      & ultrahigh      & stopping rule \&$/$ pruning enforces     & \\
 \citep{Breiman2001}   & dimensionality & minimum $n$ per terminal node & \\
 & & & \\
                                       & collinearity   & not adversely affected & \\
 & & & \\
                                       & non-linearity  & approximates well by averaging &  \\
 & &  many shallow trees & \\
                                       & interactions   &  & may describe complex interactions \\
 & & & \\
\end{tabular}
\end{sidewaystable}

\clearpage

\begin{sidewaystable}
\begin{tabular}{llll}
Method                                 & Data Characteristic & Method Pro & Method Con \\
\hline
\hline
Boruta All Relevant                    & ultrahigh      & enforces minimum $n$ per & \\
Variable Selection                    & dimensionality &  terminal node by pruning  & \\
\citep{Kursa2010}                    &                &  & \\
                                      & collinearity   &  & may erroneously exclude  \\
 & & & useful covariates \\
                                       & non-linearity  & approximates well by  & \\
 & & averaging many shallow trees & \\
 & & & \\
                                       & interactions   & may describe complex interactions & \\
 & & & \\

Boosted Trees      & ultrahigh      & stopping rule \&$/$ pruning enforces    & \\
\citep{Friedman2002}                                        & dimensionality &  minimum $n$ per terminal node & \\
 & & & \\
                                       & collinearity   & not adversely affected  & \\
 & & & \\
                                       & non-linearity  & approximates well by  & \\
 & &  averaging many shallow trees & \\
 & & & \\
                                       & interactions   & may describe complex interactions & \\
 & & & \\
\end{tabular}
\end{sidewaystable}

\clearpage

\begin{sidewaystable}
\begin{tabular}{llll}
Method              & Data Characteristic & Method Pro & Method Con \\
\hline
\hline

Cubist              & ultrahigh      & pruning inherent component of algorithm & \\
\citep{Quinlan1992} & dimensionality &                                         & \\
 & & & \\
                    & collinearity   &                                         & poorly defined coefficients   \\
 & & &  in terminal node MLRs \\
 & & & \\
                    & non-linearity  & may describe complex non-linearity & \\
 & & & \\
                    & interactions   & may describe complex interactions & \\
 & & & \\

Bayesian Treed      & ultrahigh      & tree prior tunable to pull posterior towards & computationally expensive  \\
Regression          & dimensionality & trees with sufficient $n$ per terminal node & stochastic search\\
\citep{Chipman2002} &                & for terminal node MLRs & \\
 & & & \\
                    & collinearity   &                                                         & poorly defined coefficients in terminal  \\
 & & & node MLRs, slow convergence of \\
 & & &  stochastic search \\
 & & & \\
                    & non-linearity  & Bayesian model-averaging could  & computationally expensive \\
                    &                & approximate well with shallow  & \\
                    &                & trees as could polynomial terms & \\
                    &                & in terminal node MLRs  &  \\
 & & & \\
                    & interactions   & may describe complex interactions & \\
 & & & \\
\end{tabular}
\end{sidewaystable}

\clearpage

\begin{sidewaystable}
\begin{tabular}{llll}
Method                                 & Data Characteristic & Method Pro & Method Con \\
\hline
\hline
 & & & \\
\multicolumn{4}{ l }{Abbreviations: $n$ = number of observations, MLR  = Multiple Linear Regression, CART = Classification And Regression Tree}\\
\multicolumn{4}{ l }{LASSO = Least Absolute Shrinkage and Selection Operator, LAR = Least Angle Regression}\\
 & & & \\
\hline
\hline
\end{tabular}
\end{sidewaystable}

\clearpage

\section{Appendix D: Design Matrix Filtering}
\label{App:D}
Substantial collinearity existed among the 2205 potential covariate terms.
We wished to create subsets of the full design matrix in order to explore less collinear sets of potential covariates for two reasons.
Firstly, exploring design matrices which included highly collinear pairs of covariates seemed unnecessary.
This was because a variable selection algorithm would have selected a covariate from each highly correlated pair of covariates on the basis of minute differences between the covariate values at the locations at which the response was observed and thus this decision could have been adversely influenced by errors in measurement or interpolation.
Secondly, the variable selection functions in the version of the `leaps' package \citep{Lumley2009} we used required that the design matrices explored included no pairs of covariates with correlations coefficient magnitudes greater than 0.4.
\newline
\newline
The covariates derived from the All Terrain Vehicle (ATV) surveys had the finest spatial resolution of all the covariates considered in our case study.
In an effort to build models that would have predicted the response with the greatest spatial accuracy, when faced with highly correlated pairs of covariates we chose to retain the covariates collected by the ATV surveys over any others.
Of the ATV survey derived covariates: visible Red reflectance (RED), Near InfraRed reflectance (NIR) and soil apparent electrical conductivity (ECA) only ECA had no other covariate terms derived from it while all the vegetation indices were calculated as functions of the RED and NIR reflectance values.
For this reason ECA was retained over NIR, RED and any other highly correlated covariate term.
As the vegetation indices were theoretically more indicative of green biomass than raw RED or NIR reflectance, and thus potentially more closely related to SOC levels (see \ref{App:B}), vegetation indices were retained over the raw reflectance values were any such pairs overly correlated.
Next in the order of detail of spatial resolution were the Foliar Projective Cover (FPC) Layers.
We obtained two data such layers: the projected foliage cover for 2011 (FPCI) and the projected foliage cover for 2012 (FPCII).
Since 2011 was less temporally removed from the 2009 soil survey than 2012, FPCI was set to be preferentially retained over FPCII or any other highly correlated covariates.
The coarsest spatial resolution data were derived from the Digital Elevation Model (DEM).
These data included elevation along with terrain and soil hydrology metrics calculated from the elevation.
We considered that these terrain and soil hydrology metrics came closer to describing landscape processes that may have influenced SOC formation, mineralization and or transport and thus the spatial distribution of SOC levels.
Subsequently, we elected to retain terrain and hydrology metrics over elevation should elevation have been highly correlated with any of these metrics.
Any remaining pairs of highly correlated covariates were then chosen between at random.
Once this hierarchy of filtering operations had been applied to the 63 potential covariate terms the remainder was expanded to include all remaining covariate terms to polynomial order four and all possible interactions between pairs of linear terms for these remaining covariates.
In the spirit of Occam's razor when searching the expanded design matrix for correlated pairs of covariate terms, single term polynomial terms were set to be retained in preference to any interaction terms with which they were found to be highly correlated.
Finally, lower order polynomial terms were set to be retained in preference to any higher order polynomial terms with which they were highly correlated.
Once all the above heuristics had been implemented in the order described a selection was made from any remaining pairs of highly correlated covariates at random to complete the enforcement of a maximum permitted correlation coefficient magnitude between covariates in the filtered design matrix.

\section{Appendix E: Choice of Training Set Size and Design Matrix Filtering Austerity}
\label{App:E}
We compared the results of using three different pairs of training and validation set sizes combined with each of four different levels of austerity in the design matrix filtering repeating the variable selection and model-averaging routine for each.
The training set sizes compared were 35, 45 and 55.
Sets of 500 unique training sets of each of these sizes were constructed from design matrices filtered to enforce maximum correlation coefficient magnitudes between remaining covariate pairs of 0.95, 0.8, 0.6 and 0.4.
Given our primary objective of interpolating the response between the soil core observations our focus was on out of sample predictive accuracy.
The metrics for out of sample prediction accuracy we adopted were the summary statistics for the distributions of the Validation Set Element Prediction Error (VSEPE) absolute values.
The distributions of the absolute values of the VSEPE and the coefficients of determination for the model-averaged predictions from models selected from each of these combinations of training set size and design matrix filtering austerity have been summarized in Table \ref{tab:MAP.R2.TSSxDMF}.
For each level of austerity in filtering the design matrix the distribution of the absolute values of the VSEPE appeared to become more compressed towards zero with increasing training set size.
However, it should be noted that the number of validation set elements that were predicted under the scenarios where 500 training sets of 35 observations were used was much larger compared to the number predicted under scenarios where 500 training sets of 55 observations were used ($500*(60-35)$ compared too $500*(60-55)$).
As such we only recommend comparing the VSEPE distributions obtained from each of the four collections of 500 training sets of the same size that were constructed from design matrices subjected to different the levels of filtering austerity considered.
\newline
\newline
The final model for each training set was selected from the sequence of models for that training set returned by the LAR algorithm as the model which maximised the predictive accuracy on the associated validation set.
As the 35 observation training sets were the smallest of the three sizes of training sets considered, the models selected for these training sets would have had out of sample predictive accuracy most emphasised in this second stage of their selection process where the shrinkage parameter was selected to minimise validation set predictive error.
Since areal interpolation of the response from full cover covariate observations via these models selected by LAR is in essence out of sample prediction, we elected to use the results of variable selection on one of the collections of 500 unique 35 observation training sets for this areal interpolation.
Of the models selected from the collections of 500 training sets of 35 observations it was those constructed from the design matrix filtered to enforce a maximum permitted correlation coefficient magnitude between covariate pairs of 0.95 that had the first three quarters of the ordered VSEPE absolute values most compressed towards zero.
Furthermore, of the scenarios involving 35 observation training sets it was the model-averaged prediction from the models fitted to the 500 training sets constructed from the design matrix filtered to enforce a maximum permitted correlation coefficient between covariates of 0.95 that had the best coefficient of determination.
For these reasons, we elected to interpolate the response across the study area with the model-averaged predictions from the models selected by applying LAR to the 35 observation training sets constructed from the design matrix filtered to enforce a maximum permitted correlation coefficient magnitude between remaining covariate pairs of 0.95.

\begin{table}
\caption{Summary statistics for the distributions of the absolute values of the validation set element prediction errors and coefficients of determination for model-averaged predictions from each application of Least Angle Regression to a scenario defined by the combination of training set size and design matrix filtering austerity.} 
\label{tab:MAP.R2.TSSxDMF}
\begin{tabular}{lllllllll}
  \hline
       &       & \multicolumn{6}{ c }{VSEPE} & MAP \\
$>|r|$& TSS & Min.      & 1st Qu.& Median & Mean   & 3rd Qu.& Max.  & R$^2$ \\ 
  \hline
0.95 & 35 & 1.332e-05 & 0.1482 & 0.3184 & 0.4744 & 0.5446 & 4.437 & 0.5963 \\ 
0.95 & 45 & 5.418e-05 & 0.1384 & 0.3067 & 0.4667 & 0.5410 & 4.163 & 0.6797 \\ 
0.95 & 55 & 3.734e-05 & 0.1113 & 0.2656 & 0.4065 & 0.4525 & 3.858 & 0.8403 \\ 
0.8  & 35 & 5.571e-05 & 0.1526 & 0.3288 & 0.4835 & 0.5636 & 4.167 & 0.4667 \\ 
0.8  & 45 & 5.746e-05 & 0.1462 & 0.3204 & 0.4762 & 0.5583 & 4.112 & 0.5046 \\ 
0.8  & 55 & 0.0002129 & 0.1163 & 0.2916 & 0.4220 & 0.5026 & 3.885 & 0.6284 \\ 
0.6  & 35 & 1.119e-05 & 0.1495 & 0.3362 & 0.4957 & 0.5769 & 4.206 & 0.2796 \\ 
0.6  & 45 & 0.0002723 & 0.1527 & 0.3426 & 0.5065 & 0.5916 & 4.184 & 0.3844 \\ 
0.6  & 55 & 3.376e-05 & 0.1300 & 0.2914 & 0.4193 & 0.5232 & 3.875 & 0.4994 \\ 
0.4  & 35 & 1.097e-05 & 0.1517 & 0.3324 & 0.4776 & 0.5695 & 4.063 & 0.3666 \\ 
0.4  & 45 & 0.0003481 & 0.1564 & 0.3333 & 0.4760 & 0.5662 & 3.819 & 0.4507 \\ 
0.4  & 55 & 7.387e-05 & 0.1101 & 0.2388 & 0.3626 & 0.4126 & 3.684 & 0.5593 \\ 
   \hline
\multicolumn{9}{ l }{Abbreviations:}\\ 
\multicolumn{9}{ l }{TSS = Training Set Size,}\\ 
\multicolumn{9}{ l }{VSEPE = Validation Set Element Prediction Error,} \\
\multicolumn{9}{ l }{MAP = Model-Averaged Prediction,}\\ 
\multicolumn{9}{ l }{$>|r|$ = the maximum absolute value of the correlation coefficient between}\\
\multicolumn{9}{ l }{covariate pairs permitted to remain in the design matrices supplied to the}\\
\multicolumn{9}{ l }{variable selection algorithms,}\\
\multicolumn{9}{ l }{$R^2$ = the coefficient of determination,}\\ 
\multicolumn{9}{ l }{Min. = Minimum,}\\
\multicolumn{9}{ l }{1st Qu. = First Quartile,}\\
\multicolumn{9}{ l }{3rd Qu. = Third Quartile,}\\ 
\multicolumn{9}{ l }{Max. = Maximum.}\\ 
\end{tabular}
\end{table}
\clearpage
\bibliography{FLM_Main_and_Supp.bib}

\end{document}